\documentclass[aip,jcp,reprint,floatfix,showpacs]{revtex4-1}

\usepackage{amssymb}
\usepackage{amsmath}
\usepackage{amsfonts}
\usepackage{graphicx}

\usepackage{enumitem}

\newcommand{\fancy}{\mathcal}

\newcommand{\FE}{\kappa}
\newcommand{\FG}{\fancy{G}}
\newcommand{\FP}{\fancy{P}}

\newcommand{\FW}{\vec{w}}

\newcommand{\lbrs}{\left[}
\newcommand{\rbrs}{\right]}

\renewcommand{\vec}[1]{\boldsymbol{#1}}
\newcommand{\vech}[1]{\hat{\vec{#1}}}

\newcommand{\beq}{\begin{eqnarray}}
\newcommand{\eeq}{\end{eqnarray}}

\newcommand{\tr}{\text{Tr}}
\newcommand{\Tr}[1]{{{\text{Tr}}\lbrs #1 \rbrs}}

\newcommand{\half}{\frac{1}{2}}

\newcommand{\rcite}[1]{Ref.~\onlinecite{#1}}

\newcommand{\rcites}[1]{Refs~\onlinecite{#1}}

\newcommand{\Hx}{\text{Hx}}

\newcommand{\crm}{\text{c}}

\newcommand{\EHx}{{\cal E}_{\Hx}}
\newcommand{\Ec}{{\cal E}_{\crm}}

\newcommand{\Ts}{{\cal T}_{s}}
\newcommand{\T}{{\cal T}}
\newcommand{\E}{{\cal E}}
\newcommand{\F}{{\cal F}}

\newcommand{\pr}{^{\prime}}

\ifdefined\vr
\renewcommand{\vr}{\vec{r}}
\else
\newcommand{\vr}{\vec{r}}
\fi

\newcommand{\vrp}{\vec{r}\pr}

\newcommand{\ket}[1]{\left|#1\right\rangle}

\newcommand{\iket}[1]{|#1\rangle}

\newcommand{\ibraketop}[3]{\langle#1|#2|#3\rangle}
\newcommand{\ibkouter}[1]{|#1\rangle\langle#1|}

\newcommand{\iout}{\ibkouter}

\newcommand{\iexpect}[1]{\langle#1\rangle}

\newcommand{\Fij}{\iexpect{\theta_i\theta_j}}

\newcommand{\Ext}{{\text{Ext}}}
\newcommand{\DD}{{\text{DD}}}
\newcommand{\SD}{{\text{SD}}}

\newcommand{\EXX}{{\text{EXX}}}

\newcommand{\ts}{\text{ts}}
\renewcommand{\ss}{\text{ss}}

\newcommand{\up}{\mathord{\uparrow}}
\newcommand{\down}{\mathord{\downarrow}}

\newcommand{\nh}{\hat{n}}
\newcommand{\Th}{\hat{T}}
\renewcommand{\th}{\hat{t}}
\newcommand{\Wh}{\hat{W}}
\newcommand{\vh}{\hat{v}}

\newcommand{\Hh}{\hat{H}}

\newcommand{\Gammah}{\hat{\Gamma}}

\usepackage[normalem]{ulem}
\usepackage{color}
\usepackage{xcolor}
\definecolor{NiceMagenta}{HTML}{f032e6}

\definecolor{Mygrey}{gray}{0.80}

\def\showcomment{}

\ifdefined\showcomment
\newcommand\TGComment[1]{\textcolor{blue}{{\bf[{\em Comment}: #1]}}}
\newcommand\SPComment[1]{\textcolor{red}{{\bf[{\em Comment}: #1]}}}

\newcommand\TGResponse[1]{\textcolor{blue}{{\bf[{\em Response}: #1]}}}
\newcommand\SPResponse[1]{\textcolor{red}{{\bf[{\em Response}: #1]}}}
\else
\newcommand\TGComment[1]{}
\newcommand\SPComment[1]{}

\newcommand\TGResponse[1]{}
\newcommand\SPResponse[1]{}
\fi

\begin{document}
\title{Density driven correlations in ensemble density functional
  theory: insights from simple excitations in atoms}
\author{Tim Gould}\affiliation{Qld Micro- and Nanotechnology Centre, %
  Griffith University, Nathan, Qld 4111, Australia}
\author{Stefano Pittalis}\affiliation{CNR-Istituto Nanoscienze, Via
  Campi 213A, I-41125 Modena, Italy}

\begin{abstract}
  Ensemble density functional theory extends the usual Kohn-Sham
  machinery to quantum state ensembles involving ground- and excited
  states. Recent work by the authors
  [{\em Phys. Rev. Lett.} 119, 243001 (2017); 123, 016401 (2019)]
  has shown that both the Hartree-exchange and correlation energies
  can attain unusual features in ensembles. Density-driven
  (DD) correlations -- which account for the fact that pure-state
  densities in Kohn-Sham ensembles do not necessarily reproduce those
  of interacting pure states -- are one such feature.  Here we study atoms
  (specifically $S$--$P$ and $S$--$S$ transitions) and show that the
  magnitude and behaviour of DD correlations can vary
  greatly with the variation of the orbital angular momentum of the
  involved states. Such estimations are obtained through an
  approximation for DD correlations built from relevant exact
  conditions Kohn-Sham inversion, and plausible assumptions for weakly correlated systems.
\end{abstract}

\maketitle

\section{Introduction}

In the 55 years since the Hohenberg-Kohn theorem\cite{HohenbergKohn},
it is fair to say that density functional theory (DFT) has transformed
how we study the many-electron problem.  The impact of DFT is felt far
beyond the realms of theory papers, as increasing numbers of papers
use DFT results to support experiments, to elucidate the properties of
electronic ground states. The good balance between ease-of-calculation
and accuracy of Kohn-Sham\cite{KohnSham} (KS) DFT has allowed access
to solutions to a large number of problems in chemistry and physics.

Excited states play an increasing role in chemistry\cite{Matsika2018},
however, whether via cavity modes, light--matter interactions
or otherwise. The most popular methods for calculating
reasonable excitation properties are both based on DFT, being
$\Delta$SCF and time-dependent DFT\cite{RungeGross} (TDDFT), with the
latter performed at the level of the so-called adiabatic
approximation.
The $\Delta$SCF approach is very low cost, as it involves computing
energy differences through regular self-consistent DFT calculations. 
But, it is limited to excitations between states lowest in energy and
of different symmetry species -- unless difficult orthogonality
contstaints are introduced.
TDDFT (in its conventional adiabatic sense)
is generally more robust. But, it is also more expensive than
DFT and has issues with double and charge transfer excitations.%
\cite{Maitra2005,Elliott2011,Maitra2017-CT}
Both methods are prone to spin contamination.%
\cite{Baker1993,Wittbrodt1996}
There is thus an urgent need for a more accurate treatment of
excitations that has a similar cost to conventional DFT, but avoids
the issues of $\Delta$SCF approaches.

Ensemble DFT (EDFT) is a highly promising solution to this problem,
that was prompted by Theophilou\cite{Theophilou1979} and further
developed by Gross-Oliveira-Kohn\cite{GOK-1,GOK-2,GOK-3}.  EDFT is,
conceptually, very similar to conventional pure state DFT. But it is
able to access the energies of many-electron eigenstates, not only the
ground state, in a formally exact%
\cite{Theophilou1979,Valone1980,Perdew1982,Lieb1983,Savin1996,%
  Ayers2006-Axiomatic,Gould2017-Limits,Gould2019-DD,Senjean2018} and,
as shown in recent work,%
\cite{Filatov1999-REKS,Franck2014,Filatov2015-Double,Filatov2016,Deur2017,Deur2019,%
  Yang2014,Pribram-Jones2014,Filatov2015-Review,Yang2017-EDFT,Gould2018-CT}
quantitatively accurate fashion. It thus offers a promising route to
more accurate, yet low cost, calculations of a very important class of
excitations.

In this work we review recent advancements in EDFT, that have lead to
a novel decomposition of the correlation energies for ensembles into
state- and density-driven contributions. This decomposition can help
overcome limitations in present approximations by adding corrections
or by developing innovative approximations. Next, we show how
approximate estimates of density-driven correlations can be done in
highly symmetry, weakly correlated, systems such as light atoms. Our
analysis reveals that density-driven correlations depend greatly on
the orbital angular momentum of the state involved in the considered
excitations.

This paper is organized as follows: The theory
is presented in Section~\ref{sec:Theory}, where we dedicate a
particular attention to symmetry preservation through equi-ensembles.
Results for simple excitations in atoms are presented in detail in
Section~\ref{sec:Results}.
Information on the numerical implementation are reported in
Section~\ref{sec:Methods}. Section~\ref{sec:Conclusions} concludes.

\section{Theory}\label{sec:Theory}

In any variant of density functional theory, the particle density
(hereafter referred to as the density)
is the primary variable.
The main difference between DFT and EDFT is that the former
considers  ground-state quantities as functionals of the density,
whereas the latter deals with a broader class of problem
via ensemble densities. Other than this change, the basic
structure -- i.e., the use of the variational principle -- 
is very similar to that of DFT. EDFT also allows a  consistent
handling of symmetries and makes non-interacting
$v$-representability of the interacting density more robust.%
\cite{Levy1982,Lieb1983}.

Let us thus recall the definition of an ensemble density matrix
(EDM). An EDM is a statistical average
of quantum states that can be written in operator form as:
\begin{align}\label{Gamma}
  \Gammah=&\sum_{\FE}w_{\FE}\iout{\FE},
\end{align}
for a set of non-negative weights $w_{\FE}\geq 0$ obeying
$\sum_{\FE}w_{\FE}=1$, and a set of orthonormal many-particle
wavefunctions $\{\iket{\FE}\}$.
We note that these ensembles are different to the
reduced density-matrix (RDM) used as the basic ``variable'' in
RDM functional theories (RDMFT).

Given a pure state, $\Psi$, the expectation values of an operator
$\hat{A}$ over the state, $A_{\Psi}$, may be denoted using the
bra and ket notation as follows:
\begin{align}
  A_{\Psi}=&\ibraketop{\Psi}{\hat{A}}{\Psi}\;.
\end{align}
Therefore, for a statistical mix of states as described in
Eq.~\eqref{Gamma}, the expectation values is obtained as
\begin{align}
  \fancy{A}_{\Gamma}=&\tr[\Gammah\hat{A}]=\sum_{\FE}w_{\FE}A_{\FE}\;;
\end{align}
where the ensemble average is emphasized by using calligraphic
characters. The most important measurable for our purposes
is the Hamiltonian $\Hh$, for which,
\begin{align}
  \E_{\Gamma}=&\tr[\Gammah\Hh]=\sum_{\FE}w_{\FE}E_{\FE},
\end{align}
is the ensemble average energy of the system, where
$E_{\FE}=\ibraketop{\FE}{\Hh}{\FE}$ is the energy of each state.
When no confusion may arise, we shall drop the
index $\E \equiv \E_{\Gamma}$.

For the purpose of this work we restrict EDMs to have two
additional conditions:
\begin{itemize}
\item Passive: these states can yield no work under unitary
  operations.\cite{Perarnau-Llobet2015} Mathematically,
  this means that ensemble weights obey $w_{\FE}\leq w_{\FE'}$
  when $E_{\FE}\geq E_{\FE'}$.
  Note, this condition will sometimes be partially, but rigorously,
  lifted.
\item (Weak)-equi-: ensemble weights are equal when states are
  degenerate {because of underlying symmetries of the interacting system}.
  Mathematically this means that
  $w_{\FE}=w_{\FE'}$ when $E_{\FE}=E_{\FE'}$. 
  We shall not consider equi-ensembles formed by states which
  are not degenerate, unless specifically noted. We also
  assume that ``accidental degeneracies''
  (where $\iket{\FE}$ and $\iket{\FE'}$ are not related by
  symmetry operations) are not present.
\end{itemize}
The next two section will discuss the use of the aforementioned
types of ensembles in the context of ensemble DFT for excited states.

\subsection{Ensemble DFT for excited states}\label{sec:EDFT}

Gross-Oliveira-Kohn (GOK) derived a series of
proofs\cite{GOK-1,GOK-2,GOK-3} that form the foundation of
ensemble DFT (EDFT).
A consequence of GOK's work is that we can write an energy
density functional that finds $\E$ through a variational principle
based on the ensemble density:
\begin{align}
 \E^{\FW} =&
  \min_{n}\big\{ \F^{\FW,1}[n] + \int n(\vr)v_{\Ext}(\vr) d\vr \big\}
  \label{eqn:EEDFT}
  \\
  = & \sum_{\FE}w_{\FE}E_{\FE} \;.
\end{align}
The set of ensemble weights $\FW=\{w_{\FE}\}$ form an additional input,
on top of the external potential $v_{\Ext}$. The energy is thus a
function of $\FW$ and a functional of $v_{\Ext}$.

Eq.~\eqref{eqn:EEDFT} involves energies
$E_{\FE}=\ibraketop{\FE}{\Hh}{\FE}$ which are eigenvalues of
$\Hh=\Th+\Wh+\vh_{\Ext}$ for state $\iket{\FE}$. The density
$n^{\FW}=\tr[\Gammah\nh]=\sum_{\FE}w_{\FE}n_{\FE}$ that minimizes
\eqref{eqn:EEDFT} is the density of
$\Gammah=\sum_{\FE}w_{\FE}\iout{\FE}$. Note that $\Gammah$ is passive,
so that the smallest eigenvalues are always associated with the
largest weights.

The universal functional $\F^{\FW,1}[n]$ represents the special
$\lambda=1$ case of:
\begin{align}
  \F^{\FW,\lambda}[n]=&\min_{\Gammah\to n,\FW}\Tr{
    \Gammah(\Th+\lambda\Wh) }
  \nonumber\\
  \equiv& \Tr{ \Gammah^{n^{\FW},\lambda}(\Th+\lambda\Wh) },
  \label{eqn:Flambda}
\end{align}
where $\Gammah\to n,\FW$ means $\tr[\Gammah\nh(\vr)]=n(\vr)$
and the weights of $\Gammah$ are given by $\FW$. Here,
$\tr[\Gammah^{n^{\FW},\lambda}\nh]\equiv n^{\FW}$. Except when
clarification is required, we shall now proceed to drop explicit
mention of $\FW$ on functionals and EDMs.

Furthermore, the GOK theorems guarantee that there is a unique
mapping $n\to v^{\lambda}[n]$ for any given set of ensemble
weights $\FW$ -- at least for ``well-behaved''
densities which we shall assume throughout. Thus, assuming
that interacting and non-interacting ensemble
$v$-representability is not a problem, we can define
\begin{align}
  \label{eqn:Gammahlambda}
  \Gammah^{\lambda}\equiv& \sum_{\FE}w_{\FE}\iout{\FE^{\lambda}},
\end{align}
where $\iket{\FE^{\lambda}}$ is the $\FE$th eigenfunction of
$\Hh^{\lambda}=\Th+\lambda\Wh+\vh^{\lambda}$. In the case of
equi-ensembles (see next section), these EDMs are unique, but this is
not necessarily so in general.

Taking $\lambda \rightarrow0$, and working with equi-ensembles
to preserve symmetries, we can identify $\Hh^{0}=\Th+\vh^0$ 
and the corresponding non-interacting states
$\iket{\FE_s}\equiv \iket{\FE^0}$.
These states are of the form of Slater-determinants or, when
necessary, a superposition thereof (more below).
They are formed on the same set of single-particle orbitals,
obeying self-consistent equations,
\begin{align}
  \big\{\th + v^0[n](\vr)\big\}\phi_i[n](\vr)
  =&\epsilon_i[n]\phi_i[n](\vr)\;,
  \label{eqn:KS}
\end{align}
Here $\th=-\half\nabla^2$ is the one-body kinetic energy operator,
{and $v^0$ is a spin-unpolarised potential. The orbitals,
  $\phi_i$, thus have the same spatial form for $\up$ and $\down$
  spin-channels, from which it follows that the ensemble density
  is given by,}
\begin{align}
  n(\vr)=\sum_{\FE}w_{\FE}n_{s,\FE}(\vr)\equiv \sum_if_i|\phi_i(\vr)|^2\;,
\end{align}
where $0\leq f_i\leq 2$ is an
average occupation factor. The factors,
  $f_i=\sum_{\FE}w_{\FE}\theta_i^{\FE}$, are derived from the orbital
expansion of the non-interacting densities,
\begin{align}
  n_{s,\FE}(\vr)=\sum_i\theta_i^{\FE}|\phi_i(\vr)|^2\;,
  \label{eqn:nsFE}
\end{align}
where $\theta_i^{\FE}\in\{0,1,2\}$ is the occupancy of orbital $i$
in $\iket{\FE_s}$. The potential
\begin{align}
  v_s[n](\vr)\equiv v^0[n](\vr)
\end{align}
is known as the (ensemble) Kohn-Sham potential.
For  {the equi-ensembles considered in this work}, $v_s[n]$ is unique.

Through plausible hypotheses on the ordering of the $\lambda$-dependent
eigenvalues for $\lambda \rightarrow 0$, we can define the
ensemble versions of the Kohn-Sham kinetic energy,
Hartree-exchange energy, and correlation energy.
Respectively, these are\cite{Gould2017-Limits}:
\begin{align}
  \Ts[n] \equiv &\F^0[n] = \sum_{\FE}w_{\FE} T[n_{s,\FE}]\;,
  \label{eqn:TsDef}
  \\
  \EHx[n] \equiv &\lim_{\eta\to 0^+}\frac{\F^{\eta}[n]-\T_s[n]}{\eta}\; = \sum_{\FE}w_{\FE}\Lambda_{\Hx,\FE}[n_{s,\FE}]\;.
  \label{eqn:EHxDef}
  \\
  \Ec[n]=&\F^1[n]-\Ts[n]-\EHx[n]\;.
  \label{eqn:EcDef}
\end{align}
Here, $T_{s,\FE}$ and $\Lambda_{\Hx,\FE}$ are orbital-dependent
kinetic and Hartree-exchange-like energy terms\cite{Gould2017-Limits}.
As long as we are concerned with the kinetic energy and density,
a restriction to single Slater-determinant may be harmless and
practical. But when evaluating the Hartree-exchange energy,
linear combination of Slater-determinants must be
admitted.\cite{Gould2017-Limits} Such a formalism maximally avoids
``ghost-interactions'',%
\cite{Brandi1980,Gidopoulos2002,Pastorczak2014,Yang2014}
which can cause problems in EDFT.

Next, turning to the correlation energy functional, we recognise that
it is more complex in EDFT than in DFT because -- besides the usual
reasons -- it has an additional term
that is zero in pure states.\cite{Gould2019-DD} The additional term is
required to account for the fact that a state $\FE$ is associated with
an interacting density $n_{\FE}\equiv n_{\FE}^1$ that \emph{differs}
from the corresponding non-interacting density $n_{s,\FE}\equiv
n_{\FE}^0$. Only in pure states or in certain ensembles are these two
densities guaranteed to be the same.

As a result, we arrive at
\begin{align}
  \Ec[n]=&\sum_{\FE}w_{\FE}\{ E_{\crm,\FE}^{\SD}[n_\FE, n]
  + E_{\crm,\FE}^{\DD}[n_\FE,n] \}
  \label{eqn:Ec}
\end{align}
where $E_{\crm,\FE}^{\SD}[n_\FE, n]$ is a state-driven (SD)
contribution -- a bifunctional -- which resembles the correlation
term in pure state DFT;
and $E_{\crm,\FE}^{\DD}[n_\FE,n]$ is a density-driven (DD) contribution
-- another  bifunctional -- that is unique to ensembles.\cite{Gould2019-DD}
Thus, it is the difference in the densities $n_{\FE}$ and
$n_{s,\FE}$ that gives rise to the DD term in ensembles.
Sometimes, it also makes the SD term difficult to pin down.
In the cases studied in the following sections the SD terms are
easily identified.

In order to formally describe the terms in Eq.~\eqref{eqn:Ec}
let us introduce, for each interacting state $\FE$ in the ensemble,
single-particle orbitals $\psi_i^{\FE}$ which obey a KS-like
equation \eqref{eqn:KS} but with a state-dependent potential
$v^{\FE}_{s}$ that ensures,
\begin{align}
  n_{\FE}(\vr)=\sum_i\theta_i^{\FE}|\psi_i^{\FE}(\vr)|^2\;
\end{align}
analogous to Eq.~\eqref{eqn:nsFE}.
In this step, restriction to single Slater-determinants is admissible.
More details are given in \rcites{Ayers2015, Gould2019-DD}.

Hence, we are ready to write
\begin{align}
  E_{c,\FE}^{\SD}[n_\FE,n] \equiv&
  \ibraketop{\FE}{\Th+\Wh}{\FE} - F_{\EXX,\FE}[\{\psi_i^{\FE}\}],
  \label{eqn:EcSD}
  \\
  E_{c,\FE}^{\DD}[n_\FE,n] \equiv&
  F_{\EXX,\FE}[\{\psi_i^{\FE}\}]-F_{\EXX,\FE}[\{\phi_i\}],
  \label{eqn:EcDD}
\end{align}
where $F_{\EXX,\FE}=T_{s,\FE}+\Lambda_{\Hx,\FE}$ 
{is the contribution of the state $\FE$ to the exact exchange (EXX) component of $\F^1$.}
Finally, we can also define
\begin{align}
  \F^1=& \sum_{\FE}w_{\FE}\{
  F_{\EXX,\FE}  + E_{\crm,\FE}^{\SD} + E_{\crm,\FE}^{\DD} \}\;,
  \label{eqn:FEEXXc}
\end{align}
as an alternative expression for the
interacting universal functional in eq.~\eqref{eqn:Flambda}
at $\lambda=1$.

As complicated as the bifunctionals
{of Eq.~\eqref{eqn:EcSD} and Eq.~\eqref{eqn:EcDD}} might appear,
in \rcite{Gould2019-DD} we have shown an approximate way to deal
with them practically.  In Sec.~\ref{sec:Results}, we shall
introduce another way which allows us to compute approximate
DD correlations  without having to work with bifunctional explicitly.

\subsection{Symmetries and equi-ensembles}\label{sec:Symmetries}

Although ensemble DFT may be  formulated for general weights, 
exploitation and fulfillment of symmetries imply some restrictions.
In order to explain this aspect and also to introduce the reader to
the procedure implemented in our numerical results,
let us start by briefly reviewing  a few important facts.

Firstly, the non-relativistic many-electron Hamiltonian does not
depend on the spin operators. Thus, trivially, it is invariant under
spin rotations and the many-electron states may be chosen as
eigenstates of $\hat{S}^2$ where $\hat{S}$ is the the total spin
operator. 

Secondly, some systems can also exhibit additional symmetry in real
space. The Hamiltonian of an isolated atom, for example, is invariant
under rotations in real space.  This is the reason why atomic
eigenstates may be chosen also as eigenstates of $\hat{L}^2$ where
$\hat{L}$ is the total angular momentum operator.  
Relative to a given nuclear configuration, molecules can be invariant
under the symmetry operations of point groups.  Crystal requires
consideration of space groups, which add (discrete) translational
invariance to the point groups.  Therefore the spatial symmetry of a
state is also specified by an irreducible representation, $\alpha$,
of a space symmetry group (which can, trivially, be the identity).

To further specify the terminology, let $\hat{H}$ be a Hamiltonian that
is invariant under the transformations $\hat{G}$ that forms a group
$\FG$. Provided that there are no ``accidental'' degeneracy, every
group of degenerate eigenstates provides an irreducible representation
({\em Irrep}) of $\FG$ with the dimension of the degeneracy.  Such a
representation is ``irreducible'' because the degenerate manifold is
an invariant under the action of $\FG$ {\em and} contains no
other invariant subspace.

Thus,  strictly, the corresponding eigenstates $\Psi^{\alpha}_l$
of $\hat{H}_{\FG}$ are not invariant. They transforms  as 
\begin{align}\label{tes}
  \hat{G}\ket{\Psi^{\alpha}_{l}} = & \sum_m D^{\alpha}_{ml} \ket{\Psi^{\alpha}_m}
\end{align}
where the upper index $\alpha$ labels the {\em Irrep} and the lower
index denotes the partners in the basis for the invariant subspace.
$D^{\alpha}_{ml} $ is a matrix that provides the actual representation of
$\hat{G}$.  States that transforms as the basis of a {\em Irrep} are
said to be symmetry-adapted, whether or not they may be the exact
states of the system Hamiltonian.

The Hamiltonian we work with is also invariant under exchange of
particles. The fermionic nature of the corresponding many-body states
can be taken into account by means of Slater determinants ({\em
  Sldet}) built from orthonormal single-particle states. These
determinants can be combined linearly to get {spin} symmetry-adapted
many-particle states {
  -- i.e., configuration state functions -- {to provide}
  basis functions through which we may represent either approximate
  or exact many-particle states \cite{MEST}.}
Both for computational and interpretational
convenience, especially when dealing with excited multiplets, it is
best to work with single-particle states that are symmetry adapted as
well.

The need for introducing symmetry-adaptation when accessing
excited states through ensemble density functional methods has been emphasized
by Theophilou (and various collaborators)~\cite{Theo95,Theo97,Theo00}.
In these works, weights were assigned equally to {\em all} the
states. Thus, in order to access the individual energies of the levels
of interest, several computations would be required which each involve
different ensemble energy functionals~\cite{Theophilou1979,Theo85}.
GOK\cite{GOK-2,GOK-3}
showed that the choice of the weights could be made quite flexible.%
\footnote{Recently, Fromager and coworkers\cite{Deur2019,Senjean2018}
  have pointed out that the flexibility of GOK can be exploited even
  further, in such a way to access single excited levels from
  calculations that need to refer only
  to a single ensemble-density-functional}
Equal weights, however, were still employed for degenerate states in
the same manifold. Use of equal weights is important because the
particle densities of $\iket{\Psi^{\alpha}_m}$ does not
necessarily transform according the same {\em Irrep} to which the
pure state belong. Therefore, invariant Kohn-Sham potentials can
be obtained by forming ensembles that are totally invariant
\emph{by construction}.

To be more explicit, let us consider the equi-ensemble for a
degenerate set of ground-states,
\begin{align}\label{tes2}
  \Gammah  = \frac{1}{d_{\alpha} }& \sum_{l=1}^{d_{\alpha}}
  \iout{\Psi^{\alpha}_{l}}
\end{align}
where $d_{\alpha}$ is the dimension (degeneracy) of $\alpha$.
Through use of
the orthogonality theorem, it becomes apparent that this ensemble is
invariant for any $\hat{G}\in\FG$.
The corresponding ensemble density, $n=\tr[\nh \Gammah]$ inherits
this invariance. Thus, the corresponding KS potential has the symmetry
of the external potential by construction. Ensembles for excited
states can be formed along similar lines. Explicit examples  are given
in the next sections for atoms.

Furthermore, because the Hamiltonian is an invariant w.r.t. $\FG$,
it transforms states from invariant subspaces to states of the
same subspaces. From this it readily follows that both the regular
variational principle and ensemble-variational principles may be
applied equally well to states which are symmetry
adapted. Equivalently, we may work with the projected Hamiltonian
\begin{align}\label{pH}
  \Hh_{\FP} = \hat{P}  \Hh \hat{P} 
\end{align}
where $\hat{P}$ is provided, essentially, by the same procedure used
to extract a symmetry-adapted basis from a set of generic states.

Among the advantages of this approach are:
\begin{enumerate}[label=(\alph*)]
\item When we are concerned with the lowest
  states of some  symmetry species, such as the lowest triplet or the
  lowest $^{1}D$ level in atoms, we do not need to consider any
  exited-states formulation of DFT\cite{Gunnarsson76};\label{it:1}
\item When we are concerned with the lowest two (or more) excited
  states of some symmetry species, such as the optical gap between
  singlet states, we can ignore  all the states in
  between\cite{Theo00,Gido02,Pribram-Jones2014};
\item We can find passive EDMs that are stationary on $\Hh_{\FP}$ and
  thus satisfy the weight ordering required by the GOK formulation
  despite other ``orthogonal'' states having lower energies.
  We use ``$\FP$-passive'' to denote such EDMs. The states
  described above in \ref{it:1} are the simplest example of
  $\FP$-passive states.
\end{enumerate}

Finally, we note that the above discussion allows linear
combinations of {\em Sldets} {(configuration state functions)}
at the non-interacting KS
level, so long as the terms involved all have the same energy and
density. This gives rise, through the formalism outlined previously by
the authors\cite{Gould2017-Limits}, to non-vanishing,
spontaneously emergent singlet-triplet splitting already at
the Hartree-exchange level (mentioned briefly in the previous
section).
For the purpose of the present work, however, we
average over such spin splittings and, thus, can consistently
reduce to  single {\em Sldets}, as detailed later.

\section{Results}\label{sec:Results}

Results on density-driven correlations for ensembles including
different spin states were reported in our previous work. Thus, we
focus here on quantities in ensembles that include states
that belong to different spatial {\em Irreps}. For simplicity, we
specialize our considerations to light atoms (Li, Be, Na and Mg).  
As motivated, and explained in detail, below, we exploited the
fact that mixing singlet and triplets
equally lets us to reduce the calculation to single {\em Sldets}
consistently.  

As a formal key result, we show how an approximation for the
density-driven correlation may be derived from exact conditions and
compelling assumptions that involve $s$ and $p$ orbitals.

\subsection{$S$--$P$ transitions and density-driven correlations}

Specifically, our cases involve an
excitation between $2s$--$2p$ (Li/Be) or $3s$--$3p$ (Na/Mg)
orbitals. Such transitions
are very interesting, per the results of the previous section,
as they allow calculations to be carried out across all weights,
not just passive states. This is possible because the limit $W=1$
is $\FP$-passive and thus amenable to an EDFT treatment.
Thus we can connect (via the ensemble) states that belong to
different {\em Irreps}. Importantly, the states at $W=0$ {\em and} $W=1$
may be regarded, effectively, as being ground states.

In order to illustrate the EDMs we work with, let us provide
the details for the non-interacting ones,
\begin{align}
  \Gammah_s^W=&(1-W)\Gammah_{s,S_0}+W\Gammah_{s,P_0}\;,
  \label{eqn:GammaW}
\end{align}
Here, $S_0$ refers to the equi-ensemble of states that have angular
momentum $L=0$, which can be singlet or a degenerate doublet.
$P_0$ refers to the equi-ensemble of states that have have angular
momentum $L=1$, which can be singlet or triplets.
The index ``0'' refers to the fact that we are restricting to
the lowest states in energy for each $L$. As a result, we
can reduce to work with ensemble of single {\em Sldets}.

In Eq.~(\ref{eqn:GammaW}), the first term involves an average
over a (degenerate) doublet for Li:
\begin{align}
  \Gammah_{s,S_0}^{\text{Li}}=
  \half\sum_{\sigma\in\up,\down}\iout{2s^{\sigma}}\;;
\end{align}
and is a singlet for Be:
\begin{align}
  \Gammah_{s,S_0}^{\text{Be}}=\iout{2s^2}\;.
  \label{eqn:GammaBeS0}
\end{align}
The second term involves averaging over three
{real- (or, equivalently, complex-) valued} $p$-orbitals.
It also involves averaging over a spin-doublet for Li:
\begin{align}
  \Gammah_{s,P_0}^{\text{Li}}=\frac16
  \sum_{\mu=x,y,z}\sum_{\sigma\in\up,\down}\iout{2p_{\mu}^{\sigma}}\;;
\end{align}
and equally averaging over the singlet plus triplet for Be:
\begin{align}
  \Gammah_{s,P_0}^{\text{Be}}=\frac1{12}
  \sum_{\mu=x,y,z}\sum_{\sigma,\sigma'\in\up,\down}
  \iout{2s^{\sigma} 2p_{\mu}^{\sigma'}}\;.
  \label{eqn:GammaBeShort}
\end{align}
In the expressions above, the core configuration $1s^2$ is implied.
The expressions for Na and Mg can be obtained by replacing
$1s^2\to 1s^22s^2$, $2s \to 3s$, and $2p \to 3p$.
For brevity the notation does not emphasize that the
non-interacting states also depend on $W$ (interacting quantities,
of course, do \emph{not} depend on $W$).

Finally note an important point: in the expressions above, we have
also used the fact that given a singlet-state (ss) and three triplet
states (ts$_{-1}$, ts$_{0}$ and ts$_{1}$) involving any pair of
single-particle orbitals, we can write (in short-hand)
$\frac14\big[\iout{\ss}+\sum_{s_z}\iout{\ts_{s_z}}\big]
=\frac14\sum_{\sigma,\sigma'\in\up,\down}\iout{\sigma\sigma'}$, where
the latter are the four {\em Sldets} formed on the same set of
spin-restricted spatial orbitals.  Working with such averaged states
is acceptable here, because we are interested in excitations
behaviours between different spatial {\em Irreps},
rather than excitation behaviours between different spin states.

It then follows from the properties of ensembles that the interacting
energy,
\begin{align}
  \E^W\equiv \E^W[n^W]=&(1-W)\E_{S_0}+W\E_{P_0},
  \label{eqn:EW}
\end{align}
and the density,
\begin{align}
  n^W(r)=&(1-W)n_{S_0}(r)+Wn_{P_0}(r)\;,
\end{align}
are piece-wise linear.
Note, all densities depend only on $r=|\vr|$ due to preservation
of fundamental symmetries by equi-ensembles. Here, we reintroduce
a weight superscript, $W$, to make explicit that we are referring
to an ensemble like eq.~\eqref{eqn:GammaW} taken with excitation
weights $\FW=\{1-W, W\}$ and equi-ensembles, per previous paragraphs.

Next, we {consider} the exact exchange (EXX) {contribution to the universal functional}
\begin{align}
  \F_{\EXX}^W[n^W]
  =&\F^{1,W}[n^W]-\Ec^W[n^W]
  \\
  =&\Ts^W[n^W]+\EHx^W[n^W]
  \\
  \equiv&(1-W)F_{\EXX,S_0}[\{\phi_i[n^W]\}]
  \nonumber\\&
  +WF_{\EXX,P_0}[\{\phi_i[n^W]\}]\;.
\end{align}
 {Here, the orbital functional $\F_{\EXX,S_0}$ is as defined in previous
work\cite{Gould2013-LEXX} and detailed in
  Section~\ref{sec:Methods}.} Extension to $P_0$ is described in
Section~\ref{sec:Methods}.
Note, in the final expression we use $F$, rather than
calligraphic $\F$, to indicate that the ground state ensembles
are treated like ``pure'' states. We will further clarify  this
choice later.
 
The ensemble exact exchange (EEXX) {{\em approximation}} then involves
finding,
\begin{align}
  \E_{\EXX}^W=&\min_n\big\{ \F_{\EXX}[n] + \E_{\Ext}[n] \big\}
  \nonumber\\
  \equiv& \F_{\EXX}^W[n^W_{\EXX}] + \E_{\Ext}[n^W_{\EXX}]\;,
  \label{eqn:EEXX}
\end{align}
by variational principles, using
$\E_{\Ext}[n]=\int n(\vr) v_{\Ext}(\vr)d\vr$.
Our second expression introduce the {density obtained
  self-consistently within the EEXX approximation},
$n_{\EXX}^W$, which is not the same as $n^W$.

\begin{figure}
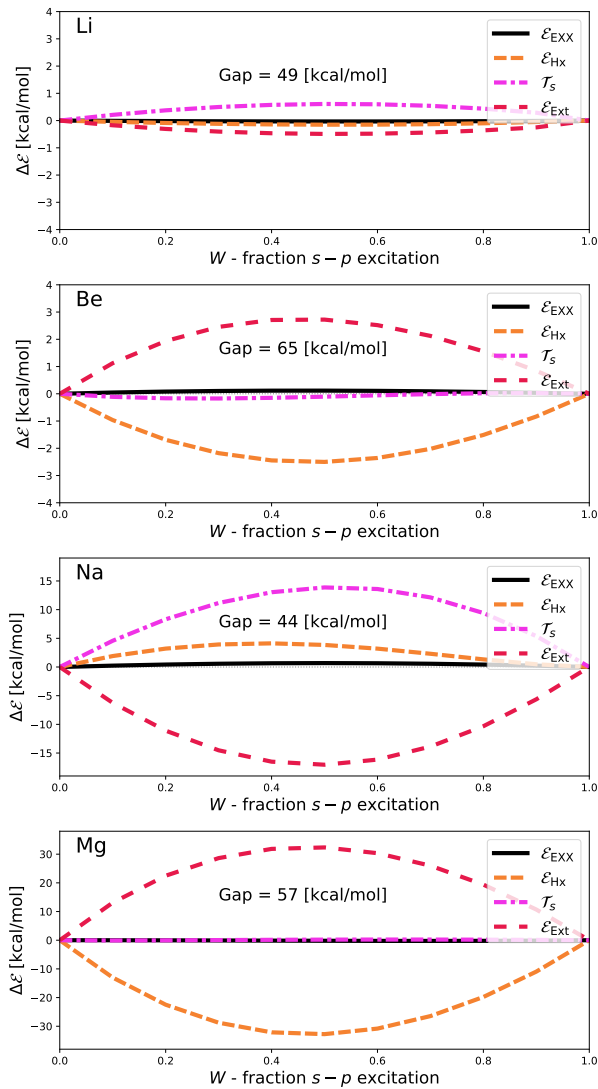

  \begin{center}
  \includegraphics[width=8cm]{{{ExciteAtom-03}}}\\
  \includegraphics[width=8cm]{{{ExciteAtom-04}}}\\
  \includegraphics[width=8cm]{{{ExciteAtom-11}}}\\
  \includegraphics[width=8cm]{{{ExciteAtom-12}}}\\
  \end{center}
  \caption{
    Non-piece-wise part $\E^W-[(1-W)\E^0+W\E^1]$ of the energy
    $\E_{\EXX}$ (solid black line) and its components
    $\EHx$ (orange dashes), $\Ts$ (magenta dashes),
    and $\E_{\text{Ext}}$ (red dashes), as a function of
    excitation weight $W$. Densities and energies obtained
    by self-consistent EXX calculations.
    Results shown for Li, Be, Na and Mg.
    The vertical gap ${\text{Gap}}=E_{\EXX,P_0}-E_{\EXX,S_0}$ is indicated
    on the plots.
  \label{fig:EEXX}}
\end{figure}

The cases $W=0$ and $W=1$ are both $\FP$-passive, and thus may be
treated as ground states. In these cases, EXX gives densities,
\begin{align}
  n_{\EXX}^{W=0}(r)\approx&n_{S_0}(r),
  &
  n_{\EXX}^{W=1}(r)\approx&n_{P_0}(r),
\end{align}
that are close to their exact counterparts in the atomic systems
we consider. Consequently, we {can approximately} treat
$n_{\EXX}^{W=0,1}$ as the true density $n^{W=0,1}$. It then follows
from $\E=\E_{\EXX}+\Ec$ that the energies,
\begin{align}
  \E_{\EXX}^{W=0}\approx&E_{S_0}-E_{\crm,S_0}^{\SD},
  &
  \E_{\EXX}^{W=1}\approx&E_{P_0}-E_{\crm,P_0}^{\SD},
\end{align}
are correct up to SD correlation terms $E_c^{\SD}$ and small errors
from the densities. 
Here we neglect DD correlations at $W=0$ and $1$. In fact, there may
be some DD correlations associated with (actual or effective) ground
state degeneracies -- we do  not concern ourselves with these
to focus on only the DD correlations that affect excitations directly.

For the cases $0<W<1$ the situation becomes more complicated. This is
because the density $n_{\EXX}^W$ that minimises Eq.~\eqref{eqn:EEXX}
is not the right ensemble density, i.e.
\begin{align}
  n_{\EXX}^W\neq {(1-W)n_{\EXX}^{W=0} + Wn_{\EXX}^{W=1}}\;.
\end{align}
Thus, when $\E_{\EXX}^W[n_{\EXX}^W]$ is employed as an estimation of the
exact energy it has two sources of error:
\begin{enumerate}
\item It misses SD and DD correlation terms as these are not
  included in the EXX approximation;
\item It is not obtained at the correct density which introduces
  an additional source of error.
\end{enumerate}
We will discuss these points further below.

First, however, we will consider how this approach affects energy
terms. Figure~\ref{fig:EEXX} shows the deviation from piece-wise
linear,
\begin{align}
  \Delta\E^W =& \E^W - [(1-W)\E^{W=0} + W\E^{W=1}].
  \label{eqn:DEW}
\end{align}
of various energy quantities. Here, energies and densities
are obtained approximately self-consistently, by solving the
Krieger-Li-Iafrate (KLI) approximation for the self-consistent
EXX potential\cite{KLI1992}. Further details of calculations
are provided in Methodology.

As the goal of these self-consistent calculations is to minimise the
total EEXX energy [Eq.~\eqref{eqn:EEXX}] neither energies nor densities
are expected to be piece-wise linear. It is thus notable that, despite
no requirement to be piece-wise linear, the EXX energy $\E_{\EXX}$
barely deviates from piece-wise linear. Adapting arguments laid out by
Gould and Dobson\cite{Gould2013-LEXX} shows that $\Delta\E_{\EXX}^W$
should be concave, but says nothing about \emph{how} concave it should
be. In fact, a very small amount of concavity can be seen for Be and
Na. In Li and Mg the concavity is invisible to the eye. In all cases
it is tiny compared to deviations of other quantities.

\begin{figure}[ht!]
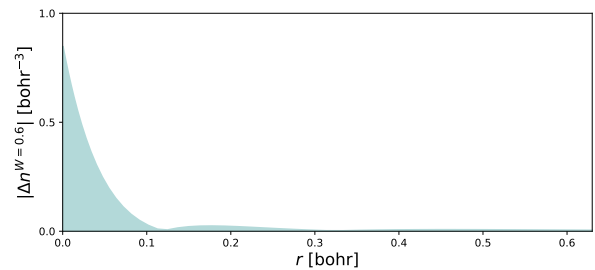

  \begin{center}
  \includegraphics[width=8cm]{{{ExciteDens-12}}}\\
  \end{center}
  \caption{
    Difference between $n_{\EXX}^W$ and piece-wise linear
    $n^W=(1-W)n_{\EXX}^{W=0}+Wn_{\EXX}^{W=1}$, for Mg.
    \label{fig:Dens}}
\end{figure}

Other energy deviations can be convex or concave, as they are only
constrained by the fact that $\E_{\EXX}$ must be concave. Indeed,
we see substantial variations in all cases except Li. In Mg, the
maximum deviation of $\EHx$ and $\E_{\Ext}$ are more than half
the vertical gap ($\text{Gap}=E_{\EXX,P_0}-E_{\EXX,S_0}$ -- equivalent to
the optical gap in atomic systems)
in magnitude, but have equal and opposite errors
which thus cancel.
Figure~\ref{fig:Dens} shows that the deviation of the density
(Mg is shown, but other atoms are qualitatively similar) is
almost exclusively found around the nucleus.

Above, we identified two sources of error found when one obtains an
energy and density using \eqref{eqn:EEXX}. We can directly remove the
second source (errors caused by deviation from piece-wise-linearity of
densities) by using ``density inversion''. That is, by seeking a
potential,
\begin{align}
  v_s[n^W]\to n^W=(1-W)n^{W=0} + Wn^{W=1},
  \label{eqn:nW}
\end{align}
that yields a set of orbials $\phi_i[n^W]$ that give our target
piece-wise linear density $n=\sum_if_i|\phi_i|^2$. This lets us
obtain $\F_{\EXX}^W[n^W]$ and $\E_{\EXX}^W[n^W]=\F_{\EXX}^W[n^W]
+\E_{\Ext}[n^W]>\E_{\EXX}^W[n_{\EXX}^W]$, and thus remove any
error caused by deviations from piece-wise linearity of the
density. Note, the external energy
$\E_{\Ext}[n^W]=\int d\vr n^W(\vr)v_{\Ext}$ is piece-wise linear,
unlike $\E_{\Ext}[n_{\EXX}^W]$. Remember, {in practice}
we approximate $n^{W=0,1}\approx n_{\EXX}^{W=0,1}$.

\begin{figure}[h]
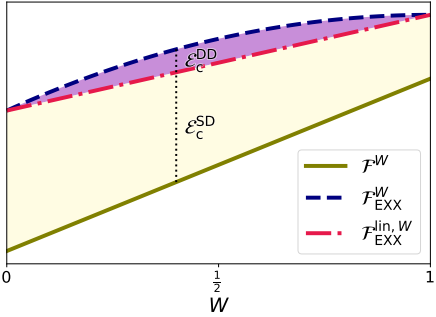

  \includegraphics[width=0.7\linewidth]{{{FigDD}}}
  \caption{
    {Qualitative} illustration of $\F^W$ (green solid line),
    $\F^{\text{lin},W}_{\EXX}$
    (red dash dot line) and $\F^W_{\EXX}$ (navy dashed line).
    Energy differences (illustrated by
    vertical black dotted lines)
    in the beige (lighter) shaded area are SD
    correlation energies. Those in the magenta
    (darker) shaded area are DD correlation energies.
    \label{fig:SDDD}}
\end{figure}

In fact, we can go further than just removing contributions from
densities. We can also use inversion to obtain a good approximation
to the DD correlation energy contribution to the excitation energy,
and thus shed light on the first identified issue. The key elements of
this approach are illustrated in Figure~\ref{fig:SDDD}

Consider the following relationships obeyed by the \emph{exact}
universal functional of our systems:
\begin{align}
  \F^W[n^W]=&\F_{\EXX}^W[n^W] + \Ec^W[n^W]
  \\
  \equiv& (1-W)F_{S_0}[n_{S_0}] +WF_{P_0}[n_{P_0}]\;.
\end{align}
Next, define the linearised (lin) EXX:
\begin{align}
  \F^{\text{lin},W}_{\EXX}[n^W]=&(1-W)\F_{\EXX}^0[n^{0}]
  + W\F_{\EXX}^{1}[n^{1}]
  \\
  \equiv& (1-W)F_{\EXX,S_0} +WF_{\EXX,P_0}\;,
\end{align}
where $F_{\EXX,\FE}=E_{\EXX,\FE}-E_{\Ext,\FE}$ for $\kappa\in S_0,P_0$.
We remind the reader we use $F$ rather than $\F$ to indicate
that the ensembles specified by the index are treated as ``pure'' states. 
The convention can now be understood in the sense that DD correlations
within (real or effective) degenerate ground-states
are treated as fixed throughout the excitation. Their
contributions may therefore be ignored.
We introduce such an assumption for the
purpose of directly focusing on the more important DD correlations in
excitations.

By definition, the SD correlation terms are,
$E^{\SD}_{c,\FE}=F^1[n_{\FE}]-F_{\EXX}[n_{\FE}]$.
It thus follows that,
\begin{align}
  \Ec^{\SD,W}[n^W] =&\F^W[n^W]
  - \F^{\text{lin},W}_{{\EXX}}[n^W]\;,
  \\
  { \approx} & {\F^W[n_{\EXX}^W]
  - \F^{\text{lin},W}_{\EXX}[n_{\EXX}^W]\;,}
\end{align}
at least as far as excitations are concerned.
{This relationship is exact if we use the interacting density
  $n^W$ for $W=0,1$ or approximate if we use $n^W_{\EXX}$.}
For arbitrary $W$ we can also write,
\begin{align}
  \F_{\EXX}^W[n^W] {\approx}&(1-W)F_{\EXX,S_0}[n_{s,S_0}^W]
  \nonumber\\&
  + WF_{\EXX,P_0}[n_{s,P_0}^W],
\end{align}
where $n^W=(1-W)n_{s,S_0}^W+Wn_{s,P_0}^W=(1-W)n_{S_0}+Wn_{P_0}$.
Here, $n_{s,S_0}$ and $n_{s,P_0}$ are formed on the same set
of orbitals from the KS potential $v_s[n^W]$ found by inversion
of $n^W$. {Here and henceforth, approximately equals signs
  indicate errors caused by treating the EXX densities for $W=0$ and
  $W=1$ as exact.  For brevity we also drop the EXX subscript on
  densities $n^{W=0}$ and $n^{W=1}$}.

Finally, using $\Ec=\Ec^{\SD}+\Ec^{\DD}=\F-\F_{\EXX}$ and
$\Ec^{\SD}=\F-\F^{\text{lin}}_{\EXX}$, we obtain
\begin{align}
  \Ec^{\DD,W}[n^W] {\approx}&\F^{\text{lin},W}_{\EXX}[n^W]-\F^W_{\EXX}[n^W]\;,
  \label{eqn:EcDDW}
\end{align}
for the DD correlation energy associated with excitations.
Thus, if we calculate EXX energies at $W=0$ and
$W=1$ and interpolate, this is approximately equivalent to
removing SD energies at all $W$. Furthermore, deviations
\begin{align}
  \Delta\Ts^W {\approx}& \Ts[n^W]-\{(1-W)\Ts[n^{W=0}]+W\Ts[n^{W=1}]\},
  \\
  \Delta\EHx^W {\approx}& \EHx[n^W]-\{(1-W)\EHx[n^{W=0}]+W\EHx[n^{W=1}]\},
\end{align}
have an equal and opposite component in the DD correlation
terms. We are consequently able to estimate
$\Ec^{\DD}$ and its components without needing to know
$\Ec^{\SD}$ -- albeit, under the assumption that
$n_{\EXX}^{W=0,1}\approx n^{W=0,1}$.

\begin{figure}[hbt]
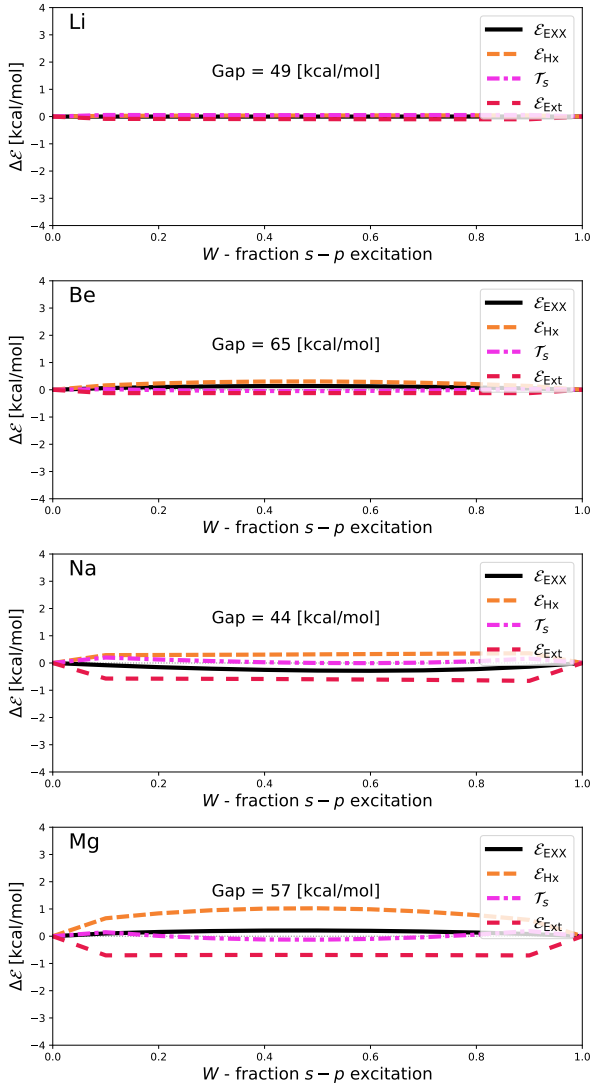

  \begin{center}
  \includegraphics[width=8cm]{{{ExciteAtom-03_Inv}}}\\
  \includegraphics[width=8cm]{{{ExciteAtom-04_Inv}}}\\
  \includegraphics[width=8cm]{{{ExciteAtom-11_Inv}}}\\
  \includegraphics[width=8cm]{{{ExciteAtom-12_Inv}}}\\
  \end{center}
  \caption{
    Like Figure~\ref{fig:EEXX} except the densities are found via
    inversion, not self-consistently.
    The black curve is approximately the negative of the
    density-driven correlation energy.
    {The vertical gap ${\text{Gap}}=E_{\EXX,P_0}-E_{\EXX,S_0}$ is indicated
    on the plots.}
  \label{fig:EInv}}
\end{figure}

Figure~\ref{fig:EInv} {reveals
that the density-driven
correlation energy is very small in our cases, because}
$\E$ barely budges
from piece-wise linearity, and is well within the margin of
errors ($\pm 1$~kcal/mol) from density inversion (here,
estimated by considering the deviation of $\E_{\Ext}$ which
is theoretically constrained to be zero). In the
case of $S$--$P$ transitions within atoms it seems that
density-driven correlation energies are likely to be small.
This bodes well for calculations of excitations obtained
from simple EDFT approximations.

\subsection{$S$--$S$ transitions and density-driven correlations}

\begin{figure}[ht!]
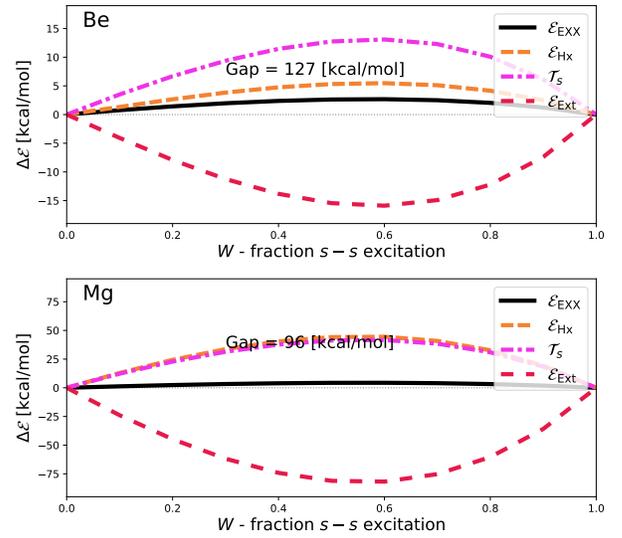

  \begin{center}
    \includegraphics[width=8cm]{{{ExciteAtom-04_ss}}}\\
    \includegraphics[width=8cm]{{{ExciteAtom-12_ss}}}\\
  \end{center}
  \caption{
    Like Figure~\ref{fig:EEXX} except showing $S$-$S$ excitations
    of Be and Mg. Energies and densities found self-consistently.
    The vertical gap ${\text{Gap}}=E_{\EXX,S_1}-E_{\EXX,S_0}$ is indicated
    on the plots.
    \label{fig:EssEXX}
  }
\end{figure}
\begin{figure}[ht!]
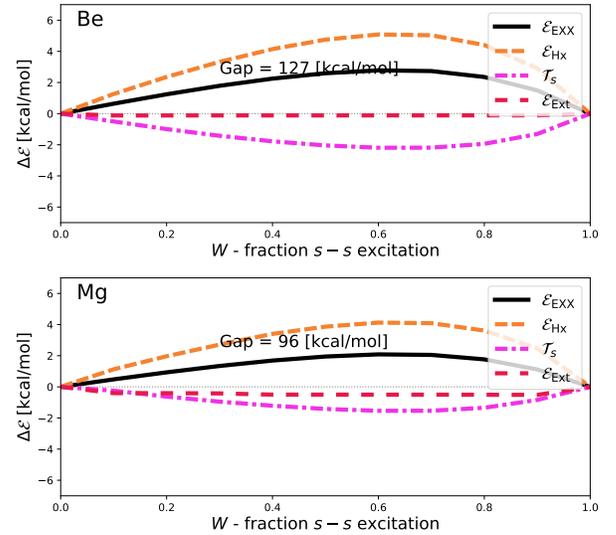

  \begin{center}
    \includegraphics[width=8cm]{{{ExciteAtom-04_ss_Inv}}}\\
    \includegraphics[width=8cm]{{{ExciteAtom-12_ss_Inv}}}\\
  \end{center}
  \caption{
    Like Figure~\ref{fig:EEXX} except showing $S$-$S$ excitations
    of Be and Mg. 
    Energies and densities found by inversion.
    The black curve is approximately the negative of the
    density-driven correlation energy.
{The vertical gap ${\text{Gap}}=E_{\EXX,S_1}-E_{\EXX,S_0}$ is indicated on the plots.}
    \label{fig:Ess}}
\end{figure}

Finally, we recognise that the $S$--$P$ transitions so far considered
are not the only ones we have available by using ensembles composed
of two ``members''. The results of
Section~\ref{sec:Symmetries} mean we can also study the
lowest lying transition between $S$ states, as such ensembles are
$\FP$-passive.

Focusing on non-interacting states for illustration, let us detail the
form of our EDMs. The full EDM is,
\begin{align}
  \Gammah_s^W\equiv&(1-W)\Gammah_{s,S_0} + W\Gammah_{s,S_1},
\end{align}
where $S_0$ is as before and $S_1$ now refers to the fact that
we average over first excited states with $L=0$.
Specifically, for Be, we study the excitation involving promotion of
the $2s$ orbital to $3s$. $\Gammah_{s,S_0}$ is defined
in Eq.~\eqref{eqn:GammaBeS0}. For $S_1$,  averaging over
singlet-triplet splitting, we get,
\begin{align}
  \Gammah^{\text{Be}}_{s,S_1}=&\frac14\sum_{\sigma,\sigma'\in\up,\down}
  \iout{2s^{\sigma} 3s^{\sigma'}}\;.
  \label{eqn:GammaBeS1}
\end{align}
For Mg we set $1s^2\to 1s^22s^2$, $2s\to 3s$ and $3s\to 4s$.

Formally, the excitations we consider can only be studied for
$0\leq W\leq \half$. However, we found that self-consistently
evaluating the fully excited state ($W=1$ -- also $\half<W\leq 1$)
worked, despite not being formally guaranteed.

Figure~\ref{fig:EssEXX} shows results obtained
self-consistently for the $2s$--$3s$ transition in Be
and the $3s$--$4s$ transition in Mg. Energies
have deviations that are, compared to the total transition
energy, very similar to the $S$--$P$ transition shown in
Figure~\ref{fig:EEXX}. However, the total EXX energy is noticeably
more concave.

Figure~\ref{fig:Ess} further highlights the concavity, and thus
the importance of DD correlations.
Unlike the cases shown in Figure~\ref{fig:EInv}, which show little
deviation, the energy $\E^W$ deviates
from piece-wise linear by up to $3$~kcal/mol for Be and
$2$~kcal/mol for Mg. As discussed above, this deviation is a
decent approximation to $\Ec^{\DD}$, which is consequently around
2.5\% of the total excitation energy of Be,
and 2\% of Mg. This is not a result of
poor inversion either, as $\E_{\Ext}$ shows minimal deviations
from piece-wise linear.

\section{Methodology}\label{sec:Methods}

Calculations are carried out in the {\tt pyAtom} Python3 code
based on numpy/scipy (available on request).
This code implements DFT in spherical
geometries. Kohn-Sham orbitals and energies of the atoms
are constructed using a spherically-symmetric {and spin-unpolarised}
Kohn-Sham potential produced by applying the
KLI\cite{KLI1992} approximation to the Greens' function,
to find the {potential, $v_{\Hx}(r)=\frac{\delta\EHx}{\delta n(r)}$,
  associated with the appropriately averaged EEXX
  energy, $\EHx$. Note $v_{\Hx}$ is a multiplicative function
  of radius $r$, not a non-local operator like in Hartree-Fock theory.}

{The primary difference from a typical EXX DFT calculation
  occurs in the definition of the Hx energy. 
  The  Hx energy of a closed-shell state is,
  $E_{\Hx}=\half\sum_{ij\in\text{occ}}[4(ii|jj)-2(ij|ji)]$,
  written in terms of the usual Coulomb integral
  $(ij|kl)=\int\frac{d\vr d\vrp}{|\vr-\vrp|}\allowbreak
  \phi_i^*(\vr)\phi_j(\vr)\phi^*_k(\vrp)\phi_l(\vrp)$.
  In EEXX,
  \begin{align}
    \EHx=&\half\sum_{ij}\big[
      (\Fij^S+\Fij^D)(ii|jj)-\Fij^S(ij|ji)
    \big]\;,
    \label{eqn:EHxEns}
  \end{align}
  has the same general form (at least in the simplified
  version we consider in this work restricting to single
  Slater determinants) but
  involves ensemble averaged pair-occupation factors,
  \begin{align}
    \Fij^S\equiv&\sum_{\FE}w_{\FE}
        [\theta_{i\up}^{\FE}\theta_{j\up}^{\FE}
          +\theta_{i\down}^{\FE}\theta_{j\down}^{\FE}]    \;,
    \label{eqn:FijS}
    \\
    \Fij^D\equiv&\sum_{\FE}w_{\FE}
        [\theta_{i\up}^{\FE}\theta_{j\down}^{\FE}
          +\theta_{i\down}^{\FE}\theta_{j\up}^{\FE}]     \;,
    \label{eqn:FijD}
  \end{align}
  for same (S) and different (D) spin pairs, respectively. Here,
  $\theta^{\FE}_{i\sigma}$ is the occupation factor for orbital $i$ with
  spin $\sigma$ in ensemble member $\FE$,
  per previous work\cite{Gould2013-LEXX,Gould2019-Hx}
  [see, especially, eq.~29 of Ref.~\onlinecite{Gould2019-Hx}].
  Eq.~\eqref{eqn:EHxEns} can thus accommodate systems with even
  and odd numbers of electrons, and high spatial
  symmetries\cite{Gould2013-LEXX}.
}

{Understanding this approach is most easily done by using an
  example: here, Li in its ground state.
  Li has a two-fold degenerate ground state which leads to a two
  member ensemble: $\FE\in\{1s^22s^{\up},1s^22s^{\down}\}$. Thus,
  we have weights $w_{1s^22s^{\up}}=w_{1s^22s^{\down}}=\half$;
  and occupation factors $\theta_{1s\up}^{\FE}=\theta_{1s\down}^{\FE}=1$
  for $1s$ states in either ensemble member, $\FE$,
  $\theta_{2s\up}^{\FE=1s^22s^{\up}}=\theta_{2s\down}^{\FE=1s^22s^{\down}}=1$
  for $2s$ states, and $\theta_{i}^{\FE}=0$ otherwise.
  Eqs.~\eqref{eqn:FijS} and \eqref{eqn:FijD} yield
  $\iexpect{\theta_{1s}\theta_{1s}}^S
  =\iexpect{\theta_{1s}\theta_{1s}}^D=2$,
  $\iexpect{\theta_{1s}\theta_{2s}}^S
  =\iexpect{\theta_{1s}\theta_{2s}}^D
  =\iexpect{\theta_{2s}\theta_{1s}}^S
  =\iexpect{\theta_{2s}\theta_{1s}}^D=1$,
  $\iexpect{\theta_{2s}\theta_{2s}}^S=1$.
  The pair-occupation factor is zero otherwise.
  Computation of $\Fij_{P_0}$ or $\Fij_{S_1}$ involves changing
  average occupation factors and accounting for any
  degeneracy due to spatial symmetries. 
  To continue our example, the $P_0$ state of Li
  (six-member ensemble of $1s^22p_m^{\sigma}$ for $\sigma\in\{\up,\down\}$
  and $m\in\{-1,0,1\}$ with $w_{\FE}=\frac16$) has
  $\iexpect{\theta_{1s}\theta_{1s}}^S
  =\iexpect{\theta_{1s}\theta_{1s}}^D=2$,
  $\iexpect{\theta_{1s}\theta_{2p_m}}^S
  =\iexpect{\theta_{1s}\theta_{2p_m}}^D
  =\iexpect{\theta_{2p_m}\theta_{1s}}^S
  =\iexpect{\theta_{2p_m}\theta_{1s}}^D=\frac13$,
  $\iexpect{\theta_{2p_m}\theta_{2p_{m'}}}^S= \frac13 \delta_{mm'}$
  or zero otherwise. The $S_1$ state has
  $\iexpect{\theta_{1s}\theta_{1s}}^S
  =\iexpect{\theta_{1s}\theta_{1s}}^D=2$,
  $\iexpect{\theta_{1s}\theta_{3s}}^S
  =\iexpect{\theta_{1s}\theta_{3s}}^D
  =\iexpect{\theta_{3s}\theta_{1s}}^S
  =\iexpect{\theta_{3s}\theta_{1s}}^D=1$,
  $\iexpect{\theta_{3s}\theta_{3s}}^S=1$
  or zero.}

The work reported here then requires the
additional step of averaging pair-occupation factors found for $S_0$
and $P_0$ (or for $S_0$ and $S_1$) according to the weights $W$ and
$1-W$, so that
\begin{align}
  \Fij^{S/D}=&
  (1-W)\Fij_{S_0}^{S/D}
  +W\Fij_{P_0}^{S/D}
  \;.
\end{align}
{We are thus able to obtain $\EHx$ directly, and then use it to
  obtain the Hx potential $v_{\Hx}(r)$.}

Orbitals $\phi_i(\vr)=R_{nl}(r)Y_{lm}(\vech{r})$ are
expanded into real radial parts $R_{nl}(r)$ obeying
\begin{align}
  \bigg[ -\frac{\partial^2}{2\partial r^2}
    + \frac{l(l+1)}{2r^2}
    + v_s(r) \bigg]rR_{nl}(r)=&\epsilon_{nl}rR_{nl}(r)\;,
  \label{eqn:RadKS}
\end{align}
and complex spherical harmonics $Y_{lm}(\vech{r})$ {(note,
  however, that -- to the end of the overall averages considered in
  this work -- it does not matter whether we choose single-particle
  orbitals with real- or complex-valued spherical harmonics).
  Here, $v_s(r)=-Z/r+v_{\Hx}(r)$ is the KS potential under the
  EEXX approximation.}

{ Calculations reported here used 160 radial values on a grid
  $r_q=Ax_q/(1-x_q)$, where abscissae $x_{0\leq q<160}=(q+\half)/160$
  are distributed evenly across the unit interval $(0,1)$, and $A$ is
  chosen based on the size of the atom involved.  The Laplacian is
  discretised using a three-point pencil. Coulomb integrals are
  obtained by quadrature, using terms of form
  \begin{align}
    &u_{L;n_1n_2n_1'n_2'}=
    \sum_{ll'}C^L_{ll'}\sum_{qq'}\omega_q\omega_{q'}
    \nonumber\\&~~~~\times
    \frac{\min[r_q,r_{q'}]^L}{\max[r_q,r_{q'}]^{L+1}}
    f_{n_1n_2,l,q}f_{n_1'n_2',l',q'}\;.
  \end{align}
  where the sum is over allowed combinations of $l,l'$,
  and $f_{n_1n_2,l}(r)=R_{n_1l}(r)R_{n_2l}(r)$. Here, $\omega_q$ are
  quadrature weights and $C^L_{ll'}$ are related to Clebsch-Gordon
  coefficients.} Tests using larger grids indicate that energy
differences are converged to within 0.1~kcal/mol.

The largest source of error in
our calculations is thus that from density inversion, i.e.
finding a potential $v^W\equiv v_s[n^W](r)$ so that the solutions
$R_{nl}(r)$ of \eqref{eqn:RadKS} obey
{$\sum_{nl}f_{nl}R_{nl}(r)^2=n^W$ [Eq.~\eqref{eqn:nW}]},
e.g., for $S$--$P$ in Be we seek to find $v^W$ such that
$2R_{1s}^2+(2-W)R_{2s}^2+WR_{2p}^2=(1-W)n_{S_0}+Wn_{P_0}$.
For our inversion algorithm we use the procedure outlined by
Garrick~\emph{et al}\cite{Garrick2019}.

\section{Conclusions}\label{sec:Conclusions}

In this work we have reviewed recent advancements in ensemble density
functional theory. We paid particular attention to manifolds of
degenerate states, the involved symmetries, and their preservation. 
 
Then, we introduced a method that uses exact exchange calculations to
estimate density-driven correlation energies, using Kohn-Sham density
inversion. This method was first demonstrated on $S$--$P$ transitions in
atoms (Li, Be, Na and Mg), which revealed that density-driven
correlation energies are very small in these transitions, suggesting
that a naive application of ensemble DFT is acceptable for these
cases {provided both states are treated using
  spin-unpolarised orbitals obtained from ensemble calculations}.
{Finally, by means of the same method, we investigated $S$--$S$
  transition in Be and Mg. Here, the density-driven correlation energy
  was shown to be much more important, being up to 3~kcal/mol
  [Figure~\ref{fig:Ess}].}

This is consistent with previous results on 1D
molecules\cite{Gould2018-CT}, where singlet-triplet transitions had
smaller DD terms than singlet-singlet transitions. It suggests
that density-driven correlations get enhanced in ensembles involving
states from different energy levels but of the same symmetry type.

Applying similar studies to molecular systems is a logical next
step. Also extension to periodically extended system is a very
compelling step to be taken shortly.  For the purpose of the present
work, we simplified our task by averaging over single-triplet
splittings.  In actual applications, however, they will have to be
resolved.  Work along these lines is being pursued.


\begin{thebibliography}{50}%
\makeatletter
\providecommand \@ifxundefined [1]{%
 \@ifx{#1\undefined}
}%
\providecommand \@ifnum [1]{%
 \ifnum #1\expandafter \@firstoftwo
 \else \expandafter \@secondoftwo
 \fi
}%
\providecommand \@ifx [1]{%
 \ifx #1\expandafter \@firstoftwo
 \else \expandafter \@secondoftwo
 \fi
}%
\providecommand \natexlab [1]{#1}%
\providecommand \enquote  [1]{``#1''}%
\providecommand \bibnamefont  [1]{#1}%
\providecommand \bibfnamefont [1]{#1}%
\providecommand \citenamefont [1]{#1}%
\providecommand \href@noop [0]{\@secondoftwo}%
\providecommand \href [0]{\begingroup \@sanitize@url \@href}%
\providecommand \@href[1]{\@@startlink{#1}\@@href}%
\providecommand \@@href[1]{\endgroup#1\@@endlink}%
\providecommand \@sanitize@url [0]{\catcode `\\12\catcode `\$12\catcode
  `\&12\catcode `\#12\catcode `\^12\catcode `\_12\catcode `\%12\relax}%
\providecommand \@@startlink[1]{}%
\providecommand \@@endlink[0]{}%
\providecommand \url  [0]{\begingroup\@sanitize@url \@url }%
\providecommand \@url [1]{\endgroup\@href {#1}{\urlprefix }}%
\providecommand \urlprefix  [0]{URL }%
\providecommand \Eprint [0]{\href }%
\providecommand \doibase [0]{http://dx.doi.org/}%
\providecommand \selectlanguage [0]{\@gobble}%
\providecommand \bibinfo  [0]{\@secondoftwo}%
\providecommand \bibfield  [0]{\@secondoftwo}%
\providecommand \translation [1]{[#1]}%
\providecommand \BibitemOpen [0]{}%
\providecommand \bibitemStop [0]{}%
\providecommand \bibitemNoStop [0]{.\EOS\space}%
\providecommand \EOS [0]{\spacefactor3000\relax}%
\providecommand \BibitemShut  [1]{\csname bibitem#1\endcsname}%
\let\auto@bib@innerbib\@empty
\bibitem [{\citenamefont {Hohenberg}\ and\ \citenamefont
  {Kohn}(1964)}]{HohenbergKohn}%
  \BibitemOpen
  \bibfield  {author} {\bibinfo {author} {\bibfnamefont {P.}~\bibnamefont
  {Hohenberg}}\ and\ \bibinfo {author} {\bibfnamefont {W.}~\bibnamefont
  {Kohn}},\ }\href {\doibase 10.1103/PhysRev.136.B864} {\bibfield  {journal}
  {\bibinfo  {journal} {Phys. Rev.}\ }\textbf {\bibinfo {volume} {136}},\
  \bibinfo {pages} {B864} (\bibinfo {year} {1964})}\BibitemShut {NoStop}%
\bibitem [{\citenamefont {Kohn}\ and\ \citenamefont {Sham}(1965)}]{KohnSham}%
  \BibitemOpen
  \bibfield  {author} {\bibinfo {author} {\bibfnamefont {W.}~\bibnamefont
  {Kohn}}\ and\ \bibinfo {author} {\bibfnamefont {L.~J.}\ \bibnamefont
  {Sham}},\ }\href {\doibase 10.1103/PhysRev.140.A1133} {\bibfield  {journal}
  {\bibinfo  {journal} {Phys. Rev.}\ }\textbf {\bibinfo {volume} {140}},\
  \bibinfo {pages} {A1133} (\bibinfo {year} {1965})}\BibitemShut {NoStop}%
\bibitem [{\citenamefont {Matsika}\ and\ \citenamefont
  {Krylov}(2018)}]{Matsika2018}%
  \BibitemOpen
  \bibfield  {author} {\bibinfo {author} {\bibfnamefont {S.}~\bibnamefont
  {Matsika}}\ and\ \bibinfo {author} {\bibfnamefont {A.~I.}\ \bibnamefont
  {Krylov}},\ }\href {\doibase 10.1021/acs.chemrev.8b00436} {\bibfield
  {journal} {\bibinfo  {journal} {Chemical Reviews}\ }\textbf {\bibinfo
  {volume} {118}},\ \bibinfo {pages} {6925} (\bibinfo {year} {2018})},\
  \bibinfo {note} {pMID: 30086645},\ \Eprint
  {http://arxiv.org/abs/https://doi.org/10.1021/acs.chemrev.8b00436}
  {https://doi.org/10.1021/acs.chemrev.8b00436} \BibitemShut {NoStop}%
\bibitem [{\citenamefont {Runge}\ and\ \citenamefont
  {Gross}(1984)}]{RungeGross}%
  \BibitemOpen
  \bibfield  {author} {\bibinfo {author} {\bibfnamefont {E.}~\bibnamefont
  {Runge}}\ and\ \bibinfo {author} {\bibfnamefont {E.~K.}\ \bibnamefont
  {Gross}},\ }\href@noop {} {\bibfield  {journal} {\bibinfo  {journal} {Phys.
  Rev. Lett.}\ }\textbf {\bibinfo {volume} {52}},\ \bibinfo {pages} {997}
  (\bibinfo {year} {1984})}\BibitemShut {NoStop}%
\bibitem [{\citenamefont {Maitra}(2005)}]{Maitra2005}%
  \BibitemOpen
  \bibfield  {author} {\bibinfo {author} {\bibfnamefont {N.~T.}\ \bibnamefont
  {Maitra}},\ }\href {\doibase 10.1063/1.1924599} {\bibfield  {journal}
  {\bibinfo  {journal} {J. Chem. Phys.}\ }\textbf {\bibinfo {volume} {122}},\
  \bibinfo {eid} {234104} (\bibinfo {year} {2005})}\BibitemShut {NoStop}%
\bibitem [{\citenamefont {Elliott}\ \emph {et~al.}(2011)\citenamefont
  {Elliott}, \citenamefont {Goldson}, \citenamefont {Canahui},\ and\
  \citenamefont {Maitra}}]{Elliott2011}%
  \BibitemOpen
  \bibfield  {author} {\bibinfo {author} {\bibfnamefont {P.}~\bibnamefont
  {Elliott}}, \bibinfo {author} {\bibfnamefont {S.}~\bibnamefont {Goldson}},
  \bibinfo {author} {\bibfnamefont {C.}~\bibnamefont {Canahui}}, \ and\
  \bibinfo {author} {\bibfnamefont {N.~T.}\ \bibnamefont {Maitra}},\
  }\href@noop {} {\bibfield  {journal} {\bibinfo  {journal} {Chem. Phys.}\
  }\textbf {\bibinfo {volume} {391}},\ \bibinfo {pages} {110} (\bibinfo {year}
  {2011})}\BibitemShut {NoStop}%
\bibitem [{\citenamefont {Maitra}(2017)}]{Maitra2017-CT}%
  \BibitemOpen
  \bibfield  {author} {\bibinfo {author} {\bibfnamefont {N.~T.}\ \bibnamefont
  {Maitra}},\ }\href@noop {} {\bibfield  {journal} {\bibinfo  {journal} {J.
  Phys.: Cond. Matter}\ }\textbf {\bibinfo {volume} {29}},\ \bibinfo {pages}
  {423001} (\bibinfo {year} {2017})}\BibitemShut {NoStop}%
\bibitem [{\citenamefont {Baker}, \citenamefont {Scheiner},\ and\ \citenamefont
  {Andzelm}(1993)}]{Baker1993}%
  \BibitemOpen
  \bibfield  {author} {\bibinfo {author} {\bibfnamefont {J.}~\bibnamefont
  {Baker}}, \bibinfo {author} {\bibfnamefont {A.}~\bibnamefont {Scheiner}}, \
  and\ \bibinfo {author} {\bibfnamefont {J.}~\bibnamefont {Andzelm}},\
  }\href@noop {} {\bibfield  {journal} {\bibinfo  {journal} {Chem. Phys.
  Lett.}\ }\textbf {\bibinfo {volume} {216}},\ \bibinfo {pages} {380} (\bibinfo
  {year} {1993})}\BibitemShut {NoStop}%
\bibitem [{\citenamefont {Wittbrodt}\ and\ \citenamefont
  {Schlegel}(1996)}]{Wittbrodt1996}%
  \BibitemOpen
  \bibfield  {author} {\bibinfo {author} {\bibfnamefont {J.~M.}\ \bibnamefont
  {Wittbrodt}}\ and\ \bibinfo {author} {\bibfnamefont {H.~B.}\ \bibnamefont
  {Schlegel}},\ }\href@noop {} {\bibfield  {journal} {\bibinfo  {journal} {J.
  Chem. Phys.}\ }\textbf {\bibinfo {volume} {105}},\ \bibinfo {pages} {6574}
  (\bibinfo {year} {1996})}\BibitemShut {NoStop}%
\bibitem [{\citenamefont {Theophilou}(1979)}]{Theophilou1979}%
  \BibitemOpen
  \bibfield  {author} {\bibinfo {author} {\bibfnamefont {A.~K.}\ \bibnamefont
  {Theophilou}},\ }\href {http://stacks.iop.org/0022-3719/12/i=24/a=013}
  {\bibfield  {journal} {\bibinfo  {journal} {J. Phys. C: Solid State Phys.}\
  }\textbf {\bibinfo {volume} {12}},\ \bibinfo {pages} {5419} (\bibinfo {year}
  {1979})}\BibitemShut {NoStop}%
\bibitem [{\citenamefont {Gross}, \citenamefont {Oliveira},\ and\ \citenamefont
  {Kohn}(1988{\natexlab{a}})}]{GOK-1}%
  \BibitemOpen
  \bibfield  {author} {\bibinfo {author} {\bibfnamefont {E.~K.~U.}\
  \bibnamefont {Gross}}, \bibinfo {author} {\bibfnamefont {L.~N.}\ \bibnamefont
  {Oliveira}}, \ and\ \bibinfo {author} {\bibfnamefont {W.}~\bibnamefont
  {Kohn}},\ }\href {\doibase 10.1103/PhysRevA.37.2805} {\bibfield  {journal}
  {\bibinfo  {journal} {Phys. Rev. A}\ }\textbf {\bibinfo {volume} {37}},\
  \bibinfo {pages} {2805} (\bibinfo {year} {1988}{\natexlab{a}})}\BibitemShut
  {NoStop}%
\bibitem [{\citenamefont {Gross}, \citenamefont {Oliveira},\ and\ \citenamefont
  {Kohn}(1988{\natexlab{b}})}]{GOK-2}%
  \BibitemOpen
  \bibfield  {author} {\bibinfo {author} {\bibfnamefont {E.~K.~U.}\
  \bibnamefont {Gross}}, \bibinfo {author} {\bibfnamefont {L.~N.}\ \bibnamefont
  {Oliveira}}, \ and\ \bibinfo {author} {\bibfnamefont {W.}~\bibnamefont
  {Kohn}},\ }\href {\doibase 10.1103/PhysRevA.37.2809} {\bibfield  {journal}
  {\bibinfo  {journal} {Phys. Rev. A}\ }\textbf {\bibinfo {volume} {37}},\
  \bibinfo {pages} {2809} (\bibinfo {year} {1988}{\natexlab{b}})}\BibitemShut
  {NoStop}%
\bibitem [{\citenamefont {Oliveira}, \citenamefont {Gross},\ and\ \citenamefont
  {Kohn}(1988)}]{GOK-3}%
  \BibitemOpen
  \bibfield  {author} {\bibinfo {author} {\bibfnamefont {L.~N.}\ \bibnamefont
  {Oliveira}}, \bibinfo {author} {\bibfnamefont {E.~K.~U.}\ \bibnamefont
  {Gross}}, \ and\ \bibinfo {author} {\bibfnamefont {W.}~\bibnamefont {Kohn}},\
  }\href {\doibase 10.1103/PhysRevA.37.2821} {\bibfield  {journal} {\bibinfo
  {journal} {Phys. Rev. A}\ }\textbf {\bibinfo {volume} {37}},\ \bibinfo
  {pages} {2821} (\bibinfo {year} {1988})}\BibitemShut {NoStop}%
\bibitem [{\citenamefont {Valone}(1980)}]{Valone1980}%
  \BibitemOpen
  \bibfield  {author} {\bibinfo {author} {\bibfnamefont {S.~M.}\ \bibnamefont
  {Valone}},\ }\href {\doibase 10.1063/1.440656} {\bibfield  {journal}
  {\bibinfo  {journal} {J. Chem. Phys.}\ }\textbf {\bibinfo {volume} {73}},\
  \bibinfo {pages} {4653} (\bibinfo {year} {1980})}\BibitemShut {NoStop}%
\bibitem [{\citenamefont {Perdew}\ \emph {et~al.}(1982)\citenamefont {Perdew},
  \citenamefont {Parr}, \citenamefont {Levy},\ and\ \citenamefont
  {Balduz}}]{Perdew1982}%
  \BibitemOpen
  \bibfield  {author} {\bibinfo {author} {\bibfnamefont {J.~P.}\ \bibnamefont
  {Perdew}}, \bibinfo {author} {\bibfnamefont {R.~G.}\ \bibnamefont {Parr}},
  \bibinfo {author} {\bibfnamefont {M.}~\bibnamefont {Levy}}, \ and\ \bibinfo
  {author} {\bibfnamefont {J.~L.}\ \bibnamefont {Balduz}},\ }\href {\doibase
  10.1103/PhysRevLett.49.1691} {\bibfield  {journal} {\bibinfo  {journal}
  {Phys. Rev. Lett.}\ }\textbf {\bibinfo {volume} {49}},\ \bibinfo {pages}
  {1691} (\bibinfo {year} {1982})}\BibitemShut {NoStop}%
\bibitem [{\citenamefont {Lieb}(1983)}]{Lieb1983}%
  \BibitemOpen
  \bibfield  {author} {\bibinfo {author} {\bibfnamefont {E.~H.}\ \bibnamefont
  {Lieb}},\ }\href {\doibase 10.1002/qua.560240302} {\bibfield  {journal}
  {\bibinfo  {journal} {Int. J. Quant. Chem.}\ }\textbf {\bibinfo {volume}
  {24}},\ \bibinfo {pages} {243} (\bibinfo {year} {1983})}\BibitemShut
  {NoStop}%
\bibitem [{\citenamefont {Savin}(1996)}]{Savin1996}%
  \BibitemOpen
  \bibfield  {author} {\bibinfo {author} {\bibfnamefont {A.}~\bibnamefont
  {Savin}},\ }\enquote {\bibinfo {title} {On degeneracy, near-degeneracy and
  density functional theory},}\ \ (\bibinfo  {publisher} {Elsevier,
  Amsterdam},\ \bibinfo {year} {1996})\ pp.\ \bibinfo {pages}
  {327--358}\BibitemShut {NoStop}%
\bibitem [{\citenamefont {Ayers}(2006)}]{Ayers2006-Axiomatic}%
  \BibitemOpen
  \bibfield  {author} {\bibinfo {author} {\bibfnamefont {P.~W.}\ \bibnamefont
  {Ayers}},\ }\href {\doibase 10.1103/PhysRevA.73.012513} {\bibfield  {journal}
  {\bibinfo  {journal} {Phys. Rev. A}\ }\textbf {\bibinfo {volume} {73}},\
  \bibinfo {pages} {012513} (\bibinfo {year} {2006})}\BibitemShut {NoStop}%
\bibitem [{\citenamefont {Gould}\ and\ \citenamefont
  {Pittalis}(2017)}]{Gould2017-Limits}%
  \BibitemOpen
  \bibfield  {author} {\bibinfo {author} {\bibfnamefont {T.}~\bibnamefont
  {Gould}}\ and\ \bibinfo {author} {\bibfnamefont {S.}~\bibnamefont
  {Pittalis}},\ }\href {\doibase 10.1103/PhysRevLett.119.243001} {\bibfield
  {journal} {\bibinfo  {journal} {Phys. Rev. Lett.}\ }\textbf {\bibinfo
  {volume} {119}},\ \bibinfo {pages} {243001} (\bibinfo {year}
  {2017})}\BibitemShut {NoStop}%
\bibitem [{\citenamefont {Gould}\ and\ \citenamefont
  {Pittalis}(2019)}]{Gould2019-DD}%
  \BibitemOpen
  \bibfield  {author} {\bibinfo {author} {\bibfnamefont {T.}~\bibnamefont
  {Gould}}\ and\ \bibinfo {author} {\bibfnamefont {S.}~\bibnamefont
  {Pittalis}},\ }\href {\doibase 10.1103/PhysRevLett.123.016401} {\bibfield
  {journal} {\bibinfo  {journal} {Phys. Rev. Lett.}\ }\textbf {\bibinfo
  {volume} {123}},\ \bibinfo {pages} {016401} (\bibinfo {year}
  {2019})}\BibitemShut {NoStop}%
\bibitem [{\citenamefont {Senjean}\ and\ \citenamefont
  {Fromager}(2018)}]{Senjean2018}%
  \BibitemOpen
  \bibfield  {author} {\bibinfo {author} {\bibfnamefont {B.}~\bibnamefont
  {Senjean}}\ and\ \bibinfo {author} {\bibfnamefont {E.}~\bibnamefont
  {Fromager}},\ }\href {\doibase 10.1103/PhysRevA.98.022513} {\bibfield
  {journal} {\bibinfo  {journal} {Phys. Rev. A}\ }\textbf {\bibinfo {volume}
  {98}},\ \bibinfo {pages} {022513} (\bibinfo {year} {2018})}\BibitemShut
  {NoStop}%
\bibitem [{\citenamefont {Filatov}\ and\ \citenamefont
  {Shaik}(1999)}]{Filatov1999-REKS}%
  \BibitemOpen
  \bibfield  {author} {\bibinfo {author} {\bibfnamefont {M.}~\bibnamefont
  {Filatov}}\ and\ \bibinfo {author} {\bibfnamefont {S.}~\bibnamefont
  {Shaik}},\ }\href {\doibase 10.1016/S0009-2614(99)00336-X} {\bibfield
  {journal} {\bibinfo  {journal} {Chem. Phys. Lett.}\ }\textbf {\bibinfo
  {volume} {304}},\ \bibinfo {pages} {429} (\bibinfo {year}
  {1999})}\BibitemShut {NoStop}%
\bibitem [{\citenamefont {Franck}\ and\ \citenamefont
  {Fromager}(2014)}]{Franck2014}%
  \BibitemOpen
  \bibfield  {author} {\bibinfo {author} {\bibfnamefont {O.}~\bibnamefont
  {Franck}}\ and\ \bibinfo {author} {\bibfnamefont {E.}~\bibnamefont
  {Fromager}},\ }\href@noop {} {\bibfield  {journal} {\bibinfo  {journal} {Mol.
  Phys.}\ }\textbf {\bibinfo {volume} {112}},\ \bibinfo {pages} {1684}
  (\bibinfo {year} {2014})}\BibitemShut {NoStop}%
\bibitem [{\citenamefont {Filatov}, \citenamefont {Huix-Rotllant},\ and\
  \citenamefont {Burghardt}(2015)}]{Filatov2015-Double}%
  \BibitemOpen
  \bibfield  {author} {\bibinfo {author} {\bibfnamefont {M.}~\bibnamefont
  {Filatov}}, \bibinfo {author} {\bibfnamefont {M.}~\bibnamefont
  {Huix-Rotllant}}, \ and\ \bibinfo {author} {\bibfnamefont {I.}~\bibnamefont
  {Burghardt}},\ }\href@noop {} {\bibfield  {journal} {\bibinfo  {journal} {J.
  Chem. Phys.}\ }\textbf {\bibinfo {volume} {142}},\ \bibinfo {pages} {184104}
  (\bibinfo {year} {2015})}\BibitemShut {NoStop}%
\bibitem [{\citenamefont {Filatov}(2016)}]{Filatov2016}%
  \BibitemOpen
  \bibfield  {author} {\bibinfo {author} {\bibfnamefont {M.}~\bibnamefont
  {Filatov}},\ }\enquote {\bibinfo {title} {Ensemble {DFT} approach to excited
  states of strongly correlated molecular systems},}\ in\ \href {\doibase
  10.1007/128_2015_630} {\emph {\bibinfo {booktitle} {Density-Functional
  Methods for Excited States}}},\ \bibinfo {editor} {edited by\ \bibinfo
  {editor} {\bibfnamefont {N.}~\bibnamefont {Ferr{\'e}}}, \bibinfo {editor}
  {\bibfnamefont {M.}~\bibnamefont {Filatov}}, \ and\ \bibinfo {editor}
  {\bibfnamefont {M.}~\bibnamefont {Huix-Rotllant}}}\ (\bibinfo  {publisher}
  {Springer International Publishing},\ \bibinfo {address} {Cham},\ \bibinfo
  {year} {2016})\ pp.\ \bibinfo {pages} {97--124}\BibitemShut {NoStop}%
\bibitem [{\citenamefont {Deur}, \citenamefont {Mazouin},\ and\ \citenamefont
  {Fromager}(2017)}]{Deur2017}%
  \BibitemOpen
  \bibfield  {author} {\bibinfo {author} {\bibfnamefont {K.}~\bibnamefont
  {Deur}}, \bibinfo {author} {\bibfnamefont {L.}~\bibnamefont {Mazouin}}, \
  and\ \bibinfo {author} {\bibfnamefont {E.}~\bibnamefont {Fromager}},\ }\href
  {\doibase 10.1103/PhysRevB.95.035120} {\bibfield  {journal} {\bibinfo
  {journal} {Phys. Rev. B}\ }\textbf {\bibinfo {volume} {95}},\ \bibinfo
  {pages} {035120} (\bibinfo {year} {2017})}\BibitemShut {NoStop}%
\bibitem [{\citenamefont {Deur}\ and\ \citenamefont
  {Fromager}(2019)}]{Deur2019}%
  \BibitemOpen
  \bibfield  {author} {\bibinfo {author} {\bibfnamefont {K.}~\bibnamefont
  {Deur}}\ and\ \bibinfo {author} {\bibfnamefont {E.}~\bibnamefont
  {Fromager}},\ }\href {\doibase 10.1063/1.5084312} {\bibfield  {journal}
  {\bibinfo  {journal} {J. Chem. Phys.}\ }\textbf {\bibinfo {volume} {150}},\
  \bibinfo {pages} {094106} (\bibinfo {year} {2019})}\BibitemShut {NoStop}%
\bibitem [{\citenamefont {Yang}\ \emph {et~al.}(2014)\citenamefont {Yang},
  \citenamefont {Trail}, \citenamefont {Pribram-Jones}, \citenamefont {Burke},
  \citenamefont {Needs},\ and\ \citenamefont {Ullrich}}]{Yang2014}%
  \BibitemOpen
  \bibfield  {author} {\bibinfo {author} {\bibfnamefont {Z.-h.}\ \bibnamefont
  {Yang}}, \bibinfo {author} {\bibfnamefont {J.~R.}\ \bibnamefont {Trail}},
  \bibinfo {author} {\bibfnamefont {A.}~\bibnamefont {Pribram-Jones}}, \bibinfo
  {author} {\bibfnamefont {K.}~\bibnamefont {Burke}}, \bibinfo {author}
  {\bibfnamefont {R.~J.}\ \bibnamefont {Needs}}, \ and\ \bibinfo {author}
  {\bibfnamefont {C.~A.}\ \bibnamefont {Ullrich}},\ }\href {\doibase
  10.1103/PhysRevA.90.042501} {\bibfield  {journal} {\bibinfo  {journal} {Phys.
  Rev. A}\ }\textbf {\bibinfo {volume} {90}},\ \bibinfo {pages} {042501}
  (\bibinfo {year} {2014})}\BibitemShut {NoStop}%
\bibitem [{\citenamefont {Pribram-Jones}\ \emph {et~al.}(2014)\citenamefont
  {Pribram-Jones}, \citenamefont {Yang}, \citenamefont {Trail}, \citenamefont
  {Burke}, \citenamefont {Needs},\ and\ \citenamefont
  {Ullrich}}]{Pribram-Jones2014}%
  \BibitemOpen
  \bibfield  {author} {\bibinfo {author} {\bibfnamefont {A.}~\bibnamefont
  {Pribram-Jones}}, \bibinfo {author} {\bibfnamefont {Z.-h.}\ \bibnamefont
  {Yang}}, \bibinfo {author} {\bibfnamefont {J.~R.}\ \bibnamefont {Trail}},
  \bibinfo {author} {\bibfnamefont {K.}~\bibnamefont {Burke}}, \bibinfo
  {author} {\bibfnamefont {R.~J.}\ \bibnamefont {Needs}}, \ and\ \bibinfo
  {author} {\bibfnamefont {C.~A.}\ \bibnamefont {Ullrich}},\ }\href {\doibase
  10.1063/1.4872255} {\bibfield  {journal} {\bibinfo  {journal} {J. Chem.
  Phys.}\ }\textbf {\bibinfo {volume} {140}},\ \bibinfo {eid} {18A541}
  (\bibinfo {year} {2014}),\ 10.1063/1.4872255}\BibitemShut {NoStop}%
\bibitem [{\citenamefont {Filatov}(2015)}]{Filatov2015-Review}%
  \BibitemOpen
  \bibfield  {author} {\bibinfo {author} {\bibfnamefont {M.}~\bibnamefont
  {Filatov}},\ }\href@noop {} {\bibfield  {journal} {\bibinfo  {journal} {WIREs
  Comput Mol Sci}\ }\textbf {\bibinfo {volume} {5}},\ \bibinfo {pages} {146}
  (\bibinfo {year} {2015})}\BibitemShut {NoStop}%
\bibitem [{\citenamefont {Yang}\ \emph {et~al.}(2017)\citenamefont {Yang},
  \citenamefont {Pribram-Jones}, \citenamefont {Burke},\ and\ \citenamefont
  {Ullrich}}]{Yang2017-EDFT}%
  \BibitemOpen
  \bibfield  {author} {\bibinfo {author} {\bibfnamefont {Z.-h.}\ \bibnamefont
  {Yang}}, \bibinfo {author} {\bibfnamefont {A.}~\bibnamefont {Pribram-Jones}},
  \bibinfo {author} {\bibfnamefont {K.}~\bibnamefont {Burke}}, \ and\ \bibinfo
  {author} {\bibfnamefont {C.~A.}\ \bibnamefont {Ullrich}},\ }\href@noop {}
  {\bibfield  {journal} {\bibinfo  {journal} {Phys. Rev. Lett.}\ }\textbf
  {\bibinfo {volume} {119}},\ \bibinfo {pages} {033003} (\bibinfo {year}
  {2017})}\BibitemShut {NoStop}%
\bibitem [{\citenamefont {Gould}, \citenamefont {Kronik},\ and\ \citenamefont
  {Pittalis}(2018)}]{Gould2018-CT}%
  \BibitemOpen
  \bibfield  {author} {\bibinfo {author} {\bibfnamefont {T.}~\bibnamefont
  {Gould}}, \bibinfo {author} {\bibfnamefont {L.}~\bibnamefont {Kronik}}, \
  and\ \bibinfo {author} {\bibfnamefont {S.}~\bibnamefont {Pittalis}},\
  }\href@noop {} {\bibfield  {journal} {\bibinfo  {journal} {J. Chem. Phys.}\
  }\textbf {\bibinfo {volume} {148}},\ \bibinfo {pages} {174101} (\bibinfo
  {year} {2018})}\BibitemShut {NoStop}%
\bibitem [{\citenamefont {Levy}(1982)}]{Levy1982}%
  \BibitemOpen
  \bibfield  {author} {\bibinfo {author} {\bibfnamefont {M.}~\bibnamefont
  {Levy}},\ }\href {\doibase 10.1103/PhysRevA.26.1200} {\bibfield  {journal}
  {\bibinfo  {journal} {Phys. Rev. A}\ }\textbf {\bibinfo {volume} {26}},\
  \bibinfo {pages} {1200} (\bibinfo {year} {1982})}\BibitemShut {NoStop}%
\bibitem [{\citenamefont {Perarnau-Llobet}\ \emph {et~al.}(2015)\citenamefont
  {Perarnau-Llobet}, \citenamefont {Hovhannisyan}, \citenamefont {Huber},
  \citenamefont {Skrzypczyk}, \citenamefont {Brunner},\ and\ \citenamefont
  {Ac\'{\i}n}}]{Perarnau-Llobet2015}%
  \BibitemOpen
  \bibfield  {author} {\bibinfo {author} {\bibfnamefont {M.}~\bibnamefont
  {Perarnau-Llobet}}, \bibinfo {author} {\bibfnamefont {K.~V.}\ \bibnamefont
  {Hovhannisyan}}, \bibinfo {author} {\bibfnamefont {M.}~\bibnamefont {Huber}},
  \bibinfo {author} {\bibfnamefont {P.}~\bibnamefont {Skrzypczyk}}, \bibinfo
  {author} {\bibfnamefont {N.}~\bibnamefont {Brunner}}, \ and\ \bibinfo
  {author} {\bibfnamefont {A.}~\bibnamefont {Ac\'{\i}n}},\ }\href {\doibase
  10.1103/PhysRevX.5.041011} {\bibfield  {journal} {\bibinfo  {journal} {Phys.
  Rev. X}\ }\textbf {\bibinfo {volume} {5}},\ \bibinfo {pages} {041011}
  (\bibinfo {year} {2015})}\BibitemShut {NoStop}%
\bibitem [{\citenamefont {Brandi}, \citenamefont {Matos},\ and\ \citenamefont
  {Ferreira}(1980)}]{Brandi1980}%
  \BibitemOpen
  \bibfield  {author} {\bibinfo {author} {\bibfnamefont {H.}~\bibnamefont
  {Brandi}}, \bibinfo {author} {\bibfnamefont {M.~D.}\ \bibnamefont {Matos}}, \
  and\ \bibinfo {author} {\bibfnamefont {R.}~\bibnamefont {Ferreira}},\ }\href
  {\doibase https://doi.org/10.1016/0009-2614(80)80726-3} {\bibfield  {journal}
  {\bibinfo  {journal} {Chem. Phys. Lett.}\ }\textbf {\bibinfo {volume} {73}},\
  \bibinfo {pages} {597 } (\bibinfo {year} {1980})}\BibitemShut {NoStop}%
\bibitem [{\citenamefont {Gidopoulos}, \citenamefont {Papaconstantinou},\ and\
  \citenamefont {Gross}(2002{\natexlab{a}})}]{Gidopoulos2002}%
  \BibitemOpen
  \bibfield  {author} {\bibinfo {author} {\bibfnamefont {N.~I.}\ \bibnamefont
  {Gidopoulos}}, \bibinfo {author} {\bibfnamefont {P.~G.}\ \bibnamefont
  {Papaconstantinou}}, \ and\ \bibinfo {author} {\bibfnamefont {E.~K.~U.}\
  \bibnamefont {Gross}},\ }\href@noop {} {\bibfield  {journal} {\bibinfo
  {journal} {Phys. Rev. Lett.}\ }\textbf {\bibinfo {volume} {88}},\ \bibinfo
  {pages} {033003} (\bibinfo {year} {2002}{\natexlab{a}})}\BibitemShut
  {NoStop}%
\bibitem [{\citenamefont {Pastorczak}\ and\ \citenamefont
  {Pernal}(2014)}]{Pastorczak2014}%
  \BibitemOpen
  \bibfield  {author} {\bibinfo {author} {\bibfnamefont {E.}~\bibnamefont
  {Pastorczak}}\ and\ \bibinfo {author} {\bibfnamefont {K.}~\bibnamefont
  {Pernal}},\ }\href {\doibase 10.1063/1.4866998} {\bibfield  {journal}
  {\bibinfo  {journal} {J. Chem. Phys.}\ }\textbf {\bibinfo {volume} {140}},\
  \bibinfo {pages} {18A514} (\bibinfo {year} {2014})},\ \Eprint
  {http://arxiv.org/abs/https://doi.org/10.1063/1.4866998}
  {https://doi.org/10.1063/1.4866998} \BibitemShut {NoStop}%
\bibitem [{\citenamefont {Ayers}, \citenamefont {Levy},\ and\ \citenamefont
  {Nagy}(2015)}]{Ayers2015}%
  \BibitemOpen
  \bibfield  {author} {\bibinfo {author} {\bibfnamefont {P.~W.}\ \bibnamefont
  {Ayers}}, \bibinfo {author} {\bibfnamefont {M.}~\bibnamefont {Levy}}, \ and\
  \bibinfo {author} {\bibfnamefont {A.}~\bibnamefont {Nagy}},\ }\href {\doibase
  10.1063/1.4934963} {\bibfield  {journal} {\bibinfo  {journal} {The Journal of
  Chemical Physics}\ }\textbf {\bibinfo {volume} {143}},\ \bibinfo {pages}
  {191101} (\bibinfo {year} {2015})},\ \Eprint
  {http://arxiv.org/abs/https://doi.org/10.1063/1.4934963}
  {https://doi.org/10.1063/1.4934963} \BibitemShut {NoStop}%
\bibitem [{\citenamefont {Helgaker}, \citenamefont {J{\o}rgensen},\ and\
  \citenamefont {Olsen}(2002)}]{MEST}%
  \BibitemOpen
  \bibfield  {author} {\bibinfo {author} {\bibfnamefont {T.}~\bibnamefont
  {Helgaker}}, \bibinfo {author} {\bibfnamefont {P.}~\bibnamefont
  {J{\o}rgensen}}, \ and\ \bibinfo {author} {\bibfnamefont {J.}~\bibnamefont
  {Olsen}},\ }\href
  {https://www.ebook.de/de/product/3344768/frank_l_pilar_elementary_quantum_chemistry_secon.html}
  {\emph {\bibinfo {title} {Molecular Electronic-Structure Theory}}}\ (\bibinfo
   {publisher} {John Wiley {\&} Sons, LTD.},\ \bibinfo {year}
  {2002})\BibitemShut {NoStop}%
\bibitem [{\citenamefont {Theophilou}\ and\ \citenamefont
  {Gidopoulos}(1995)}]{Theo95}%
  \BibitemOpen
  \bibfield  {author} {\bibinfo {author} {\bibfnamefont {A.~K.}\ \bibnamefont
  {Theophilou}}\ and\ \bibinfo {author} {\bibfnamefont {N.~I.}\ \bibnamefont
  {Gidopoulos}},\ }\href {\doibase 10.1002/qua.560560418} {\bibfield  {journal}
  {\bibinfo  {journal} {Int. J. Quantum Chem.}\ }\textbf {\bibinfo {volume}
  {56}},\ \bibinfo {pages} {333} (\bibinfo {year} {1995})}\BibitemShut
  {NoStop}%
\bibitem [{\citenamefont {Theophilou}(1997)}]{Theo97}%
  \BibitemOpen
  \bibfield  {author} {\bibinfo {author} {\bibfnamefont {A.~K.}\ \bibnamefont
  {Theophilou}},\ }\href@noop {} {\bibfield  {journal} {\bibinfo  {journal}
  {Int. J. Quantum Chem.}\ }\textbf {\bibinfo {volume} {61}},\ \bibinfo {pages}
  {333} (\bibinfo {year} {1997})}\BibitemShut {NoStop}%
\bibitem [{\citenamefont {Theophilou}\ and\ \citenamefont
  {Papaconstantinou}(2000)}]{Theo00}%
  \BibitemOpen
  \bibfield  {author} {\bibinfo {author} {\bibfnamefont {A.~K.}\ \bibnamefont
  {Theophilou}}\ and\ \bibinfo {author} {\bibfnamefont {P.~G.}\ \bibnamefont
  {Papaconstantinou}},\ }\href {\doibase 10.1103/PhysRevA.61.022502} {\bibfield
   {journal} {\bibinfo  {journal} {Phys. Rev. A}\ }\textbf {\bibinfo {volume}
  {61}},\ \bibinfo {pages} {022502} (\bibinfo {year} {2000})}\BibitemShut
  {NoStop}%
\bibitem [{\citenamefont {Hadjisavvas}\ and\ \citenamefont
  {Theophilou}(1985)}]{Theo85}%
  \BibitemOpen
  \bibfield  {author} {\bibinfo {author} {\bibfnamefont {N.}~\bibnamefont
  {Hadjisavvas}}\ and\ \bibinfo {author} {\bibfnamefont {A.}~\bibnamefont
  {Theophilou}},\ }\href {\doibase 10.1103/PhysRevA.32.720} {\bibfield
  {journal} {\bibinfo  {journal} {Phys. Rev. A}\ }\textbf {\bibinfo {volume}
  {32}},\ \bibinfo {pages} {720} (\bibinfo {year} {1985})}\BibitemShut
  {NoStop}%
\bibitem [{Note1()}]{Note1}%
  \BibitemOpen
  \bibinfo {note} {Recently, Fromager and coworkers\cite {Deur2019,Senjean2018}
  have pointed out that the flexibility of GOK can be exploited even further,
  in such a way to access single excited levels from calculations that need to
  refer only to a single ensemble-density-functional}\BibitemShut {NoStop}%
\bibitem [{\citenamefont {Gunnarsson}\ and\ \citenamefont
  {Lundqvist}(1976)}]{Gunnarsson76}%
  \BibitemOpen
  \bibfield  {author} {\bibinfo {author} {\bibfnamefont {O.}~\bibnamefont
  {Gunnarsson}}\ and\ \bibinfo {author} {\bibfnamefont {B.~I.}\ \bibnamefont
  {Lundqvist}},\ }\href {\doibase 10.1103/PhysRevB.13.4274} {\bibfield
  {journal} {\bibinfo  {journal} {Phys. Rev. B}\ }\textbf {\bibinfo {volume}
  {13}},\ \bibinfo {pages} {4274} (\bibinfo {year} {1976})}\BibitemShut
  {NoStop}%
\bibitem [{\citenamefont {Gidopoulos}, \citenamefont {Papaconstantinou},\ and\
  \citenamefont {Gross}(2002{\natexlab{b}})}]{Gido02}%
  \BibitemOpen
  \bibfield  {author} {\bibinfo {author} {\bibfnamefont {N.}~\bibnamefont
  {Gidopoulos}}, \bibinfo {author} {\bibfnamefont {P.}~\bibnamefont
  {Papaconstantinou}}, \ and\ \bibinfo {author} {\bibfnamefont
  {E.}~\bibnamefont {Gross}},\ }\href {\doibase 10.1016/S0921-4526(02)00799-8}
  {\bibfield  {journal} {\bibinfo  {journal} {Physica B}\ }\textbf {\bibinfo
  {volume} {318}},\ \bibinfo {pages} {328} (\bibinfo {year}
  {2002}{\natexlab{b}})},\ \bibinfo {note} {proceedings of the
  6\textsuperscript{th} Patras University Euroconference on Proper ties of
  Condensed Matter Probed with X-ray Scattering - Electron Correla tions and
  Magnetism}\BibitemShut {NoStop}%
\bibitem [{\citenamefont {Gould}\ and\ \citenamefont
  {Dobson}(2013)}]{Gould2013-LEXX}%
  \BibitemOpen
  \bibfield  {author} {\bibinfo {author} {\bibfnamefont {T.}~\bibnamefont
  {Gould}}\ and\ \bibinfo {author} {\bibfnamefont {J.~F.}\ \bibnamefont
  {Dobson}},\ }\href {\doibase 10.1063/1.4773284} {\bibfield  {journal}
  {\bibinfo  {journal} {J. Chem. Phys.}\ }\textbf {\bibinfo {volume} {138}},\
  \bibinfo {eid} {014103} (\bibinfo {year} {2013})}\BibitemShut {NoStop}%
\bibitem [{\citenamefont {Krieger}, \citenamefont {Li},\ and\ \citenamefont
  {Iafrate}(1992)}]{KLI1992}%
  \BibitemOpen
  \bibfield  {author} {\bibinfo {author} {\bibfnamefont {J.~B.}\ \bibnamefont
  {Krieger}}, \bibinfo {author} {\bibfnamefont {Y.}~\bibnamefont {Li}}, \ and\
  \bibinfo {author} {\bibfnamefont {G.~J.}\ \bibnamefont {Iafrate}},\
  }\href@noop {} {\bibfield  {journal} {\bibinfo  {journal} {Phys. Rev. A}\
  }\textbf {\bibinfo {volume} {45}},\ \bibinfo {pages} {101} (\bibinfo {year}
  {1992})}\BibitemShut {NoStop}%
\bibitem [{\citenamefont {Gould}\ \emph {et~al.}(2019)\citenamefont {Gould},
  \citenamefont {Pittalis}, \citenamefont {Toulouse}, \citenamefont
  {Kraisler},\ and\ \citenamefont {Kronik}}]{Gould2019-Hx}%
  \BibitemOpen
  \bibfield  {author} {\bibinfo {author} {\bibfnamefont {T.}~\bibnamefont
  {Gould}}, \bibinfo {author} {\bibfnamefont {S.}~\bibnamefont {Pittalis}},
  \bibinfo {author} {\bibfnamefont {J.}~\bibnamefont {Toulouse}}, \bibinfo
  {author} {\bibfnamefont {E.}~\bibnamefont {Kraisler}}, \ and\ \bibinfo
  {author} {\bibfnamefont {L.}~\bibnamefont {Kronik}},\ }\href {\doibase
  10.1039/C9CP03633D} {\bibfield  {journal} {\bibinfo  {journal} {Phys. Chem.
  Chem. Phys.}\ }\textbf {\bibinfo {volume} {21}},\ \bibinfo {pages} {19805}
  (\bibinfo {year} {2019})}\BibitemShut {NoStop}%
\bibitem [{\citenamefont {Garrick}\ \emph {et~al.}(2019)\citenamefont
  {Garrick}, \citenamefont {Natan}, \citenamefont {Gould},\ and\ \citenamefont
  {Kronik}}]{Garrick2019}%
  \BibitemOpen
  \bibfield  {author} {\bibinfo {author} {\bibfnamefont {R.}~\bibnamefont
  {Garrick}}, \bibinfo {author} {\bibfnamefont {A.}~\bibnamefont {Natan}},
  \bibinfo {author} {\bibfnamefont {T.}~\bibnamefont {Gould}}, \ and\ \bibinfo
  {author} {\bibfnamefont {L.}~\bibnamefont {Kronik}},\ }\href {\doibase
  10.26434/chemrxiv.8869460.v1} {\bibfield  {journal} {\bibinfo  {journal}
  {Under review}\ } (\bibinfo {year} {2019}),\
  10.26434/chemrxiv.8869460.v1}\BibitemShut {NoStop}%
\end{thebibliography}
\end{document}